\documentclass[prd,showpacs,preprintnumbers,twocolumn,amsmath,amssymb,superscriptaddress,nofootinbib,english]{revtex4-1}
\usepackage{graphicx}
\usepackage{dcolumn}
\usepackage{bm}
\usepackage{epsfig}
\usepackage{graphicx}
\usepackage{hyperref}
\usepackage[usenames]{color}
\usepackage{url}
\hypersetup{
    colorlinks=true,
    linkcolor=black,
    citecolor=black,
}

\newcommand{\rd}{{\rm d}}

\newcommand{\remove}[1]{}
\newcommand{\mpl}{m_{\rm Pl}}
\newcommand{\mpm}{m_{\rm Pl}^{-1}}

\def\be{\begin{equation}}
\def\ee{\end{equation}}
\def\ba{\begin{eqnarray}}
\def\ea{\end{eqnarray}}

\begin{document}

\title{A Unified Description of Screened Modified Gravity }

\author{Philippe~Brax}
\email[Email address: ]{philippe.brax@cea.fr}
\affiliation{Institut de Physique Theorique, CEA, IPhT, CNRS, URA 2306, F-91191Gif/Yvette Cedex, France}

\author{Anne-Christine~Davis}
\email[Email address: ]{a.c.davis@damtp.cam.ac.uk}
\affiliation{DAMTP, Centre for Mathematical Sciences, University of Cambridge, Wilberforce Road, Cambridge CB3 0WA, UK}

\author{Baojiu~Li}
\email[Email address: ]{baojiu.li@durham.ac.uk}
\affiliation{ICC, Physics Department, University of Durham, South Road, Durham DH1 3LE, UK}

\author{Hans A. Winther}
\email[Email address: ]{h.a.winther@astro.uio.no}
\affiliation{Institute of Theoretical Astrophysics, University of Oslo, 0315 Oslo, Norway}

\date{\today}

\begin{abstract}
We consider  modified gravity models driven by a scalar field whose effects are screened in high density regions due to the presence of non-linearities in its interaction potential and/or its coupling to matter. Our approach covers chameleon, $f(R)$ gravity, dilaton and symmetron models and allows a unified description of all these theories. We find that the dynamics of modified gravity are entirely captured by  the time variation of the scalar field mass and its coupling to matter evaluated at the cosmological minimum of its effective potential, where the scalar field sits since an epoch prior to Big Bang Nucleosynthesis. This new parameterisation of modified gravity allows one to reconstruct the potential and coupling to matter and therefore to analyse the full  dynamics of the models, from the scale dependent growth of structures at the linear level to non-linear effects requiring $N$-body simulations. This procedure is illustrated with  explicit examples of reconstruction for chameleon, dilaton,  $f(R)$ and symmetron models.
\end{abstract}

\maketitle

\section{Introduction}

The discovery of the acceleration of the expansion of the Universe \cite{cst2006} has led to a reappraisal of some of the tenets of modern cosmology. In particular, the possibility of modifying the laws of gravity on short or large scales is taken more and more seriously \cite{cfps2011}.

In view of Weinberg's theorem stating that any Lorentz invariant field theory involving spin-2 fields must reduce to General Relativity (GR) at low energy  \cite{weinberg}, any attempt to modify GR must involve extra degree(s) of freedom. The majority of known models involve scalar fields and can be separated into two broad classes, the ones involving non-linearities in the kinetic terms and others with non-linear interaction potentials. All these models have a coupling of the scalar field to matter and there could be an environmental dependence which would manifest itself in the screening behaviour of the scalar field in high density regions \cite{k2010,bdl2012}. Examples of such models abound: the dilatonic models \cite{dilaton,dilaton2} generalising the Damour-Polyakov mechanism \cite{dp1994} where the coupling to gravity turns off in dense environments, the chameleon models \cite{chameleon,bbdkw2004,bbd2004,bbmnw2010,gmmptw2010} where a thin shell shielding the scalar field in dense bodies is present, the symmetron models \cite{Pietroni:2005pv,Olive:2007aj,symmetron,hklm2011,bbdlss2011,dlmw2012,cjk2012} where the scalar field has a  symmetry breaking potential where the field is decoupled at high density.

Some models are essentially spin-offs of the previous ones like the $f(R)$ theories \cite{starobinsky,cdtt2004,fR,ftbm2007,nv2007,agpt2007,Carloni:2007br,shs2007,lb2007,hs2007,bbds2008} (for recent reviews of the $f(R)$ gravity see \cite{fr_review_sf,fr_review_dt}) which are only valid when they behave like chameleon theories with a thin shell mechanism in dense environments \cite{bbds2008}. In all these examples, the large scale properties on cosmological distances are intimately linked to the small scale physics as probed in the solar system or laboratory tests of gravity. Stringent constraints on the possible modifications of gravity follow from the cosmology of these models too. In particular, they may lead to potentially lethal variations of particle masses or Newton's constant during Big bang Nucleosynthesis (BBN). This must be avoided at all cost as this may destroy the formation of elements, one of the big successes of the Big Bang model. Such a catastrophe can be avoided provided the scalar fields sit at the minimum of the density dependent effective potential prior to BBN. If this is the case, then the minimum of these models is stable enough to prevent large excursions of the scalar field and therefore of scalar masses/Newton's constant when the electron decouples during BBN. One of the most important consequences of this fact, which is common to  chameleons, dilatons and symmetrons is that the cosmological background with the scalar field at the density dependent minimum of the effective potential behaves essentially like the $\Lambda$-Cold Dark Matter ($\Lambda$CDM) model and is therefore almost indistinguishable from a cosmology comprising matter, radiation and a pure cosmological constant. This is a major drawback and would immediately render irrelevant the modified gravity/dark energy models with screening properties.

Fortunately, this is far from being the case as first anticipated in \cite{bbdkw2004,Brax:2005ew} where the equation governing the density contrast of CDM was first studied. Indeed, inside the Compton wavelength of the scalar field, the density contrast grows anomalously compared to its usual growth in the matter dominated era. If this discrepancy were large enough on astrophysical scales, this may be detectable by future galaxy surveys. It turns out that the perturbation equation at the linear level depends on the time evolution of the scalar field mass and the coupling strength to matter. With these two functions, all the time and space properties of the linear perturbations can be calculated.

In fact, these two time-dependent functions capture a lot more about the modified gravity models with screening properties: they allow one to reconstruct fully and uniquely the whole non-linear dynamics of the models \cite{bdl2012,Brax:2011ta}. Hence given these two functions, not only can one compute linear perturbations, but one can study the gravitational properties of the models in the solar system and laboratory experiments. One can also analyse the cosmological behaviour of the models with $N$-body simulations. This way of defining the models, a reversed engineering procedure from the mass and coupling functions to the non-linear dynamics, is a lot more versatile than the usual direct route where a model is defined by its Lagrangian comprising the kinetic terms and an interacting potential. Indeed, all the usual models such as chameleons, $f(R)$, dilaton  and symmetrons can be explicitly rediscovered by specifying the particular ways the mass and coupling functions behave in time. Moreover, one can design new families of  models. At the linear level of cosmological perturbations, this approach is equivalent to a space and time dependent parameterisation \cite{bertschinger2006,hs2007b,jz2008,aks2008,bz2008,sk2009,bt2010,dlsccll2010,pskz2010,zllkbz2011} in terms of the two Newtonian potentials obtained in the Jordan frame: the modified Poisson equation and the constitutive relation linking the two Newtonian potentials are directly and uniquely determined by the mass and coupling functions in the Einstein frame. For instance, we shall see below that one recovers the phenomenological description of $f(R)$ models which uses a space and time dependent parameterisation \cite{bz2008} as a simple application of our formalism.

The paper is arranged as follows, in a first part we describe the modified gravity models with scalar fields and their cosmological background and gravitational properties. We then describe the tomography of the models, i.e. how to reconstruct their full dynamics using the time evolution of the mass and coupling functions. In section IV, we focus on $f(R)$ models. In section V we analyse the growth of structure. In section VI, we consider the constraints on these models resulting from the variation of the fundamental constants. We conclude in section VII.

Throughout this paper the metric convention is chosen as $(-,+,+,+)$; Greek indices ($\mu,\nu,\cdots$) run over $0,1,2,3$ while Latin indices ($i,j,k,\cdots$) run over $1,2,3$. We shall adopt the unit $c=1$ and $m_{\rm Pl}$ denotes the Planck mass. Unless otherwise stated a subscript $_0$ will always mean the present-day value of a quantity.

\section{Modified Gravity}

\label{sect:modified_gravity}

In this paper we propose a parameterisation of a broad class of theories with a scalar degree of freedom, such as the chameleon, dilaton and symmetron theories, and  $f(R)$ gravity. The success of these theories relies on  mechanisms that suppress the fifth force in  local, high matter-density, environments. We will find that the complete non-linear Lagrangian comprising the kinetic terms and the interaction potential together with the coupling of the scalar field to matter can be reconstructed from the knowledge of the scalar field mass $m(a)$ and the coupling strength $\beta (a)$ as functions of time when the field sits at the minimum of the density dependent effective potential.

This mechanism relies on the fact that the scalar field must track that minimum since before BBN in order to preserve the constancy of particle masses at this epoch. In this section, we recall the setting of scalar field models and analyse their background evolution.

\subsection{Modifying Gravity with a Scalar Field}

The action governing the dynamics of a scalar field $\phi$ in a scalar-tensor theory is
of the general form
\begin{eqnarray}\label{eq:action}
S &=& \int {\rm d}^4x\sqrt{-g}\left\{\frac{m_{\rm Pl}^2}{2}{R}-\frac{1}{2}(\nabla\phi)^2- V(\phi)\right\}\nonumber\\
&& + \int {\rm d}^4x \sqrt{-\tilde g} {\cal L}_m(\psi_m^{(i)},\tilde g_{\mu\nu}),
\end{eqnarray}
where $g$ is the determinant of the metric $g_{\mu\nu}$, ${ R}$ is the Ricci scalar and $\psi_m^{(i)}$ are various matter fields labelled by $i$. A key ingredient of the model is the conformal coupling of $\phi$ with matter particles. More precisely, the excitations of each matter field $\psi_m^{(i)}$ couple to a metric $\tilde g_{\mu\nu}$ which is related to the Einstein-frame metric $g_{\mu\nu}$ by the conformal rescaling
\be
\tilde g_{\mu\nu}=A^2(\phi)g_{\mu\nu}.
\ee
The metric $\tilde g_{\mu\nu}$ is the Jordan frame metric. We will analyse these models in the Einstein frame and come back to the Jordan frame picture later.

The fact that the scalar field couples to matter implies that the scalar field equation becomes density dependent. More precisely, the scalar field equation of motion (EOM) is modified due to the coupling of the scalar field $\phi$ to matter:
\be
\Box \phi= -\beta T + \frac{{\rm d}V}{{\rm d}\phi},
\ee
where $T$ is the trace of the energy momentum tensor $T^{\mu\nu}$, $\Box\equiv\nabla^\mu\nabla_\mu$ and the coupling of $\phi$ to matter is defined by
\be
\beta(\phi) \equiv \mpl\frac{{\rm d}\ln A}{{\rm d} \phi}.
\ee
This is equivalent to the usual scalar field EOM with the effective potential
\be
V_{\rm eff}(\phi) = V(\phi) - \left[A(\phi)-1\right]T.
\ee
The role of this effective potential $V_{\rm eff}(\phi)$ is crucial in all the modified gravity models we will consider. In essence, the effective potential is required to possess a unique matter dependent minimum in the presence of pressure-less matter where $T=-\rho_m$. The resulting potential
\be
V_{\rm eff}(\phi) = V(\phi) +[A(\phi)-1]\rho_m
\ee
has a minimum $\phi_{\rm min}(\rho_m)$. The mass of the scalar field at the minimum
\be
m^2 = \frac{\rd^2 V_{\rm eff}}{\rd\phi^2}\big|_{\phi_{\rm min}}
\ee
must be positive. In many cases (such as the generalised chameleon and dilaton models discussed below) $V(\phi)$ is a decreasing function and $\beta(\phi)$ is an increasing function as $\phi$, though this is not the case for the generalised symmetron model\footnote{For the generalised symmetron models, the potential is not monotonic but has the shape of a Mexican hat. However, in the part of the potential which will be of interest here, it is monotonically decreasing.}. This guarantees that the effective potential always has a minimum. In a cosmological setting we will also impose that $m^2\gg H^2$ with $H$ being the Hubble expansion rate. It can be shown easily that, depending on the shapes of $V(\phi)$ and $\beta (\phi)$, the chameleon, $f(R)$, dilaton and symmetron models are all described in a such a way.

When matter is described by a pressure-less fluid with
\be
T^{\mu\nu}= \rho_m u^\mu u^\nu
\ee
and $u^\mu\equiv\rd x^\mu/\rd\tau$ where $\tau$ is the proper time, the matter density $\rho_m$ is conserved
\be
\dot \rho_m +\theta\rho_m=0
\ee
where $\theta\equiv\nabla_\mu u^\mu$ and the trajectories are determined by the modified geodesics
\be
\dot u^\mu + \beta \frac{\dot \phi}{m_{\rm Pl}} u^\mu= - \beta \frac{\partial^\mu \phi}{m_{\rm Pl}}.
\ee
In the weak-field limit with
\be
\rd s^2=-(1+2\Phi_N) \rd t^2+ (1-2\Phi_N)\rd x^i\rd x_i,
\ee
and in the non-relativistic case, this reduces to the modified geodesic equation for matter particles
\be
\frac{\rd^2 x^i}{\rd t^2}= -\partial^i\left(\Phi_N+\ln A(\phi) \right).
\ee
This can be interpreted as the motion of a particle in the effective gravitational potential defined as
\be
\Psi=\Phi_N+\ln A(\phi),
\ee
and is clearly a manifestation of the dynamics of modified gravity.

When a particle of mass $M$ in a homogeneous background matter density is the source of gravity, the scalar field satisfies
\be
\left(\nabla^2+m^2\right) \phi= \beta \frac{M}{\mpl}\delta^{(3)}(r),
\ee
in which $\delta^{(3)}(r)$ is the 3-dimensional Dirac $\delta$-function and $m$ the scalar field mass in the background, implying that
\be
\Psi= -\left(1+2\beta^2e^{-mr}\right)\frac{G_N M}{r},
\ee
where $G_N=(8\pi)^{-1}m^{-2}_{\rm Pl}$ is the Newton constant. When $\beta\sim{\cal O}(1)$ and $m^{-1}\gg r$, this implies a substantial deviation from Newton's law. For bodies much bigger than a point particle following the modified geodesics, non-linear effects imply that the effective coupling felt by the body is much smaller than $\beta$ or the mass becomes much larger than the inverse of the typical size of the body ($m^{-1}\ll r$). This is what happens in the chameleon model and $f(R)$ gravity (the latter) and the dilaton and symmetron models (the former), and guarantees that solar system and laboratory tests of gravity are evaded.

\subsection{Screening of Modified Gravity}

In this section, we shall unify the description for the screening\footnote{To be clear, the 'screening' of a body refers to the fact that the deviation from Newtonian gravity, {\it i.e.,} the fifth force exerted by this body on a nearby test mass, is suppressed to evade local constraints -- in analogy to the screening of the electric force from a charged particle.} mechanisms \cite{k2010,bdl2012} involved in the chameleon, $f(R)$ gravity, dilaton and symmetron models. As we shall see, the screening of large and dense bodies can be expressed with a single criterion generalising the thin-shell condition for the chameleon models. The constraints we find are typically stated in terms of the scalar field mass $m_0$ in the cosmological background today and the current Hubble scale $H_0$, making $\xi \equiv H_0/m_0$ a key quantity. Physically, $\xi$ represents the range of the scalar fifth-force to the Hubble radius and a particular value that will be recurrent is $m_0/H_0 \sim 10^3$ or $\xi \sim 10^{-3}$. This value means that the scalar field leaves its mark up to scales of the order of mega parsec, which again signals  the transition where the modifications of gravity can be seen on linear perturbations or not.

\subsubsection{ Chameleons}

The chameleon models (at least in their original form \cite{chameleon}, see  \cite{bbdkw2004,bbd2004,bbmnw2010,gmmptw2010} for other proposals) are characterised by a runaway potential and a nearly constant coupling $\beta$. Chameleons are screened deep inside a massive body, where the field settles at the minimum $\phi_c$ of $V_{\rm eff}(\phi)$ and stays constant up until a radius $R_s$ close to the radius of the body, $R$. In this case, the field profile is given by
\be
\phi= \phi_c, \ R\le R_s
\ee
The field varies sharply inside a thin shell according to
\be
\frac{1}{r^2}\frac{\rd}{\rd r} \left[r^2 \frac{\rd\phi}{\rd r}\right]= \beta \frac{\rho_m}{\mpl},\ \  R_s\le r\le R
\ee
and decays outside
\be
\phi= \phi_\infty - \frac{\beta}{{4\pi \mpl}}\left[1-\frac{R_s^3}{R^3}\right]\frac{M}{r} \frac{e^{-m_\infty (r-R)}}{r}
\ee
where $\phi_\infty$ is the minimum of the effective potential outside the body and $m_{\infty}, M$ are respectively the masses of the scalar field and the body. At short distance compared to the large range $m_\infty^{-1}$, the effective gravitational potential is
\be
\Psi= \beta \frac{\phi_\infty }{\mpl}+ \frac{G_N M}{r}\left[1+{2\beta^2}\left(1-\frac{R_s^3}{R^3}\right)\right].
\ee
Gravity is strongly modified by a factor $(1+2\beta^2)$ if there is no shell inside the body ({\it i.e.}, $R_s=0$) and one retrieves GR when $R_s$ is close to $R$ where
\be
\frac{\Delta R}{R}= \frac{\vert\phi_\infty -\phi_c\vert}{6\beta\mpl\Phi_N},
\ee
with  $\Delta R\equiv R-R_s$ and $\Phi_N\equiv G_N M/R$ is the Newtonian potential at the surface of the body. The mass is screened when
\be
\vert \phi_\infty -\phi_c\vert\ll 2\beta\mpl\Phi_N,
\ee
which is also the criterion to have a thin shell.

More precisely, this implies several very stringent experimental constraints on the chameleon models. The first one comes from the Lunar Ranging experiment \cite{Williams:2012nc} which measures the acceleration difference between the Earth and the Moon in the gravitational field of the Sun
\be
\eta= \frac{2(a_{\rm earth}-a_{\rm moon})}{a_{\rm earth}+a_{\rm moon}}\lesssim 10^{-13}.
\ee
For the chameleon model we have \cite{chameleon}
\be
\eta \approx \beta^2\left(\frac{\Delta R_\oplus}{R_\oplus}\right)^2,
\ee
implying that
\be
\beta\frac{\Delta R_\oplus}{R_\oplus}\lesssim 10^{-7}.
\ee
The Cassini experiment \cite{Bertotti:2003rm} imposes that the modification of the unscreened Cassini satellite in the vicinity of the sun should be such that
\be
\beta^2 \frac{\Delta R_\odot}{R_\odot} \lesssim 10^{-5}.
\ee
Another type of constraint comes from cavity experiments where two small test bodies interact in a vacuum cavity \cite{eotwash}. This implies that
\be
\beta\frac{\Delta R_{\rm cav}}{R_{\rm cav}}\lesssim 10^{-3}.
\ee
Finally, a loose bound must be imposed to guarantee that galaxies are not far off from being Newtonian \cite{Pourhasan:2011sm}
\be
\beta \frac{\Delta R_{\rm gal}}{R_{\rm gal}}\lesssim 1,
\ee
otherwise the modifications of gravity would have been seen by now in observations of galaxy clusters. These constraints strongly restrict the parameter space of the chameleon models.

\subsubsection{Symmetrons}

Symmetrons \cite{symmetron,hklm2011,bbdlss2011,dlmw2012,cjk2012} are models with a mexican hat potential, a local maximum at the origin and two global minima at $\pm\phi_\star$ like for example
 \be
V(\phi)= V_0  + \mu^2\phi_\star^2\left[-\frac{1}{2}\left(\frac{\phi}{\phi_\star}\right)^2 + \frac{1}{4}\left(\frac{\phi}{\phi_\star}\right)^4\right].
\ee
In general the term $(\phi/\phi_\star)^4$ can be replaced by any even function which is bounded below, without changing the qualitative properties of the model.

Meanwhile, the coupling behaves like
\be
A(\phi)= 1+ \frac{A_2}{2} \phi^2,
\ee
close to $\phi=0$.

Let us consider a spherically dense body that is embedded in a homogeneous background. Inside this body the matter density $\rho_m$ is constant and the scalar field profile is
\be
\phi= C \frac{\sinh m_cr}{r},~~~r<R,
\ee
where the scalar field mass is given by $m^2= A_2 \rho_m -\mu^2$ and $-\mu^2$ is the negative curvature of the potential $V(\phi)$ at the origin. The field outside the body, on scales shorter than the large range $m_\infty^{-1}$ associated to the scalar field value $\phi_\infty$ which minimises $V_{\rm eff}(\phi)$ outside, is
\be
\phi= \phi_\infty + \frac{D}{r},~~~r>R,
\ee
where
\begin{eqnarray}
C &=& \frac{\phi_\infty}{m_c\cosh m_cR},\nonumber\\
D &=& \frac{ \sinh m_cR -m_cR \cosh m_cR}{m_c\cosh m_cR}\phi_\infty.
\end{eqnarray}
If the body is dense enough, we have $m_c^2 \approx A_2 \rho_m$ and $m_cR \gg 1$, implying that $D\approx -R\phi_\infty$. Identifying the coupling to matter $\beta_\infty = \mpl A_2 \phi_\infty$, we find that the modified Newtonian potential outside the body is
\begin{eqnarray}
\Psi &=& -\frac{G_N M}{r}\left[1+\frac{A_2\phi_\infty^2}{\Phi_N}\right]+{\cal O}(\frac{R^2}{r^2})\nonumber\\
&=&  -\frac{G_N M}{r}\left[1+\frac{\beta_\infty^2}{A_2\mpl^2 \Phi_N}\right]+{\cal O}(\frac{R^2}{r^2}).
\end{eqnarray}
for $r$ sufficiently large compared to $R$.   For $R\ll r\ll m_\infty^{-1}$
the fifth-force is screened provided
\be
2A_2\mpl^2 \Phi_N \gg 1,
\ee
which is equivalent to
\be
\vert \phi_\infty -\phi_c \vert \ll 2\mpl\beta_\infty \Phi_N,
\ee
where $\phi_c=0$. Note that this is the same screening criterion as in the chameleon case.

The screening in the symmetron model depends on $A_2$, $\Phi_N$ and the environment through the environmental field value $\phi_{\infty}$. Two test masses which are not screened when put in vacuum will be screened by a factor $(\phi_{\infty}/\phi_\star)^2$ if they are in a region of high matter density (which implies $\phi_{\infty} \ll \phi_\star$).

The transition of the minimum of $V_{\rm eff}(\phi)$ from $\phi=0$ to $\phi=\phi_\star$ in the cosmological background happens in the recent past of the Universe provided
\be
\mu^2 \sim A_2 \rho_{m0},
\ee
where $\rho_{m0}$ is the present matter density. For a polynomial potential $V(\phi)$, the mass-squared $m^2_\star$  at the minimum $\phi_\star$ is of order $\mu^2$, implying that the mass of symmetrons in the present cosmological background satisfies
\be
m_0^2 \sim A_2\mpl^2  H_0^2,
\ee
One may see effects of modified gravity on astrophysical scales when $m_0/H_0 \lesssim 10^3$ which implies that $A_2 \mpl^2 \lesssim 10^6$.

Using the screening criterion we find that the Sun and the Milky Way with $\Phi_\odot \sim 10^{-6}$ are marginally screened whereas the Earth with $\Phi_\oplus \sim 10^{-9}$ and the Moon with $\Phi_{\rm moon} \sim 10^{-11}$ are not screened. However, for the solar system tests such as the Lunar Ranging experiment\footnote{The Nordtvedt effect leads to a weak bound \cite{symmetron}.}  and the Cassini satellite, what is more relevant is the value of the symmetron field $\phi_{\rm gal}$ in the Milky Way, which determines the strength $\beta(\phi_{\rm gal})$ of the modification of gravity.

This  imposes
\be
\frac{A_2\phi_{\rm gal}^2}{\Phi_\odot}\lesssim 10^{-5}.
\ee
For a generic symmetron potential we have\footnote{See Eq.~(19) in \cite{symmetron} for a more accurate expression.} $\phi_{\rm gal}^2 \sim \frac{\rho_{\infty}}{\rho_{\rm gal}}\phi_\star^2$ where $\phi_\star$ is the minimum of $V_{\rm eff}(\phi)$ in the cosmological background with matter density $\rho_\infty$. Using $\frac{\rho_\infty}{\rho_{\rm gal}}\sim 10^{-6}$, this leads to
\be
10^{-6} \beta_\star^2\frac{1}{2A_2\mpl^2 \Phi_\odot}\sim \frac{10^{-6}}{\Phi_\odot}\frac{H_0^2}{m_0^2} \lesssim 10^{-5}
\ee
which is easily satisfied for $m_0/H_0\sim 10^3$. Finally, in cavity experiments, the field $\phi$ inside the cavity is almost identical to the field in the bore, {\it i.e.}, $\phi\sim 0$, implying no deviation from usual gravity in such experiments.

\subsubsection{Dilaton}

Dilatonic theories \cite{dilaton,dilaton2} are very similar to symmetrons in as much as they share the same type of coupling function,
\be
A(\phi)=1+ \frac{A_2}{2}(\phi-\phi_\star)^2,
\ee
but they differ as the dilaton potential $V(\phi)$ is a monotonically decreasing function of $\phi$. All the dynamics can be analysed in the vicinity of $\phi_\star$ as the minimum of the effective potential is close to $\phi_\star$ for large enough $A_2$.

The density dependent minimum of $V_{\rm eff}(\phi)$ is given by
\be
\phi_{\rm min}(\rho_m)-\phi_\star = -\frac{V'(\phi_\star)}{A_2 \rho_m},
\ee
with the mass given by
\be
m^2= m^2_\star +A_2 \rho_m,
\ee
where $m_\star=m(\phi_\star)$ and the potential is chosen to be a quintessence potential such that $m^2_\star \sim H_0^2$.

Let us consider a spherically dense body. Inside the body we have
\be
\phi= \phi_c + C \frac{\sinh m_c r}{r}, \ r<R,
\ee
and outside
\be
\phi= \phi_\infty + \frac{D}{r},
\ee
for distances shorter than the range $m_\infty^{-1}$. When $m_cR\gg 1$, we find that
\be
D\approx -R (\phi_\infty -\phi_c),
\ee and
the effective Newtonian potential is
\be
\Psi=-\frac{G_N M}{r} \left[1 + \frac{A_2 (\phi_\infty -\phi_c)(\phi_\infty -\phi_\star)}{\Phi_N}\right]+{\cal O}(\frac{R^2}{r^2}),
\ee
for $ R \ll r\ll m_{\infty}^{-1}$.
Outside the body we have
\be
\phi_\infty -\phi_\star= \frac{\beta_\infty}{A_2 \mpl}
\ee
with $\beta_\infty=\beta(\phi_\infty)$ and therefore
\be
V'(\phi_\star)=-\beta_\infty \frac{\rho_\infty}{\mpl},
\ee
from which we deduce that
\be
\phi_\infty -\phi_c= \frac{\beta_\infty}{A_2 \mpl}\left(1-\frac{\rho_\infty}{\rho_c}\right),
\ee
and finally
\be
\Psi= -\frac{G_N M}{r} \left[1+ \frac{\beta^2_\infty}{A_2 \mpl^2\Phi_N}\left(1-\frac{\rho_\infty}{\rho_c}\right)\right]+{\cal O}(\frac{R^2}{r^2}).
\ee
for $ R \ll r\ll m_{\infty}^{-1}$.
The screening criterion is (almost) the same as in the symmetron case
\be
2A_2 \mpl^2\Phi_N\gg \left(1-\frac{\rho_\infty}{\rho_c}\right),
\ee
or equivalently
\be
\vert \phi_\infty -\phi_c\vert \ll 2 \beta (\phi_\infty)\mpl \Phi_N,
\ee
which is the same as in the chameleon and dilaton cases.

The mass of the dilaton today in the cosmological background is
\be
m^2_0 \approx A_2 \rho_{m0} = 3 A_2 \mpl^2 \Omega_{m0}H_0^2,
\ee
in which $\Omega_{m0}$ is the present value of the fractional energy density of matter $\Omega_m$, implying that $A_2\mpl^2 \sim 10^6$ for models with $m_0/H_0 \sim 10^3$.

As in the symmetron case, this implies that both the sun and the Milky Way are marginally screened when surrounded by the cosmological vacuum. But given that what matters for the magnitude of modified gravity is the dilaton value $\phi_\infty=\phi_{\rm gal}$ in the Milky Way, the Cassini bound can be written as
\be
\frac{A_2 (\phi_{\rm gal} -\phi_c)(\phi_{\rm gal} -\phi_\star)}{\Phi_N}\lesssim 10^{-5},
\ee
which leads to
\be
\frac{1}{A_2 \mpl^2\Phi_\odot} \frac{\rho_\infty}{\rho_{\rm gal}}\lesssim 10^{-5}.
\ee
Using $\frac{\rho_\infty}{\rho_{\rm gal}}\sim 10^{-6}$, we see that the Cassini bound is satisfied for dilatons.

\subsubsection{The Screening Criterion}

We have seen that all the  models of the chameleon, dilaton and symmetron types lead to a screening mechanism provided that
\be
\vert \phi_\infty -\phi_c\vert \ll 2 \beta (\phi_\infty)\mpl \Phi_N,
\ee
where $\phi_c$ is the value inside the body assumed to be at the minimum of the effective potential, $\phi_\infty$ the minimum value outside the body and $\Phi_N$ is Newton's potential at the surface of the body. This is a universal criterion which is {\it independent} of the details of the model. In fact, it depends only on the values of the scalar field which minimises the effective potential $V_{\rm eff}(\phi)$ inside and outside the body.  If this criterion is satisfied, then the value inside the body does not deviate much from the minimum value there.

Phenomenologically, we have just recalled that stringent local constraints on modified gravity can be expressed in terms of the screening condition. In the following we shall {\it assume that the Milky Way satisfies the screening criterion}. When this is the case, local tests of gravity in the solar system and in the laboratory can be easily analysed as $\phi_{\rm gal}$ can be determined analytically. In the chameleon, dilaton and symmetron cases, this allows one to determine bounds on the ratio $m_0/H_0$ which essentially dictates if modified gravity has effects on astrophysical scales. The screening condition for the Milky Way may be relaxed slightly for some model parameters because it is itself in a cluster with higher density than the background. In this case, full numerical simulations are required to determine $\phi_{\rm gal}$ and see if local tests of gravity are satisfied. This may enlarge the allowed parameter space of the models slightly and lead to interesting effects. Numerical simulations are left for future work.

One of the advantages of the screening condition is that it only depends on the minimum values of the scalar field in different matter densities.
In the following section, we will find an explicit formula for $\phi_c -\phi_\infty$ which depends only on the time variation of the mass $m(a)$ and coupling $\beta(a)$ in a cosmological background. This may seem surprising as the behaviour of the scalar field may appear to be loosely connected to the scalar field dynamics in a static environment. In fact, the relation between both regimes of modified gravity, cosmological and static, follows from the fact that  the scalar field sits at the minimum of its effective potential $V_{\rm eff}(\phi)$ since before BBN. As it evolves from BBN through the dark ages and then the present epoch, the cosmological values of the scalar field experience all the possible minima of $V_{\rm eff}(\phi)$. Hence realising a tomography of the cosmological behaviour of the scalar field, {\it i.e.}, just knowing its mass and coupling to matter as a function of time since before BBN, will allow us to analyse the gravitational properties of the models.

\subsubsection{The Reason for a Universal Screening Condition}

As we have seen in the examples above, we get the same screening condition for all known models. Below we argue why this is the case for a whole range of models satisfying only some simple assumptions.

We start with the most general model for the behaviour of the scalar field in matter
\begin{align}
\nabla^2\phi = V_{\rm eff,\phi} = V_{,\phi} + \frac{\beta(\phi)\rho_m}{\mpl}
\end{align}
and we will analyse the standard setup -- a spherical body of density $\rho_c$ and radius $R$ embedded in a background of density $\rho_{\infty}$ -- under the following assumptions:
\begin{enumerate}
\item The effective potential has a matter dependent minimum $\phi(\rho)$.
\item For any (physical) solution to the field equation, the mass of the field at $r=0$, $m_S = m(\phi_S,\rho_c)$, is a positive monotonically increasing function of the density $\rho_c$ and satisfies\footnote{As $\rho_c\to\infty$ we have $\phi_S\to\phi_c$; the minimum for the matter density $\rho_c$. The reason we explicitly write the limit here instead of taking $\phi_S = \phi_c$ directly is to account for models where $\lim_{\phi\to \phi_c}V_{\rm eff,\phi\phi} = 0$, but where  $\lim_{\rho_c \to \infty}V_{\rm eff,\phi\phi}(\phi_S(\rho_c),\rho_c) = \infty$ as can be the case for generalised symmetron models as we shall see later on. Loosely speaking we can state this condition as: the mass at the minimum inside the body is increasing with $\rho_c$.} $\lim_{\rho_c \to \infty} m(\phi_S(\rho_c),\rho_c) = \infty$.
\item Outside the body, where $\rho_{\infty} \ll \rho_c$, and within the Compton wavelength of the field $m_{\infty}^{-1}$ the solution to the field equation is well approximated by $\phi = \phi_{\infty} + \frac{D}{r}$. This means that a first order Taylor expansion around $\phi_{\infty}$ holds outside the body.
\end{enumerate}
Now we can look at the solutions to the field equation under the previous assumptions. The field starts out at some field-value $\phi = \phi_S$ inside the body, and close to $r=0$ the solution can therefore be written
\begin{eqnarray}
\phi = \phi_S + B\left(\frac{\sinh(m_S r)}{m_S r}-1\right)
\end{eqnarray}
for some constant $B$. We can for our purposes, without loss of generality, assume that $B>0$. Because of our assumption on $m_S$, for a large enough $\rho_c$ the field must start off very close to the minimum $\phi = \phi_c$ inside the body where the driving force $V_{\rm eff,\phi}$ vanishes. Otherwise the solution ($\sim e^{m_S r}/r$) grows too fast inside the body and overshoots the exterior solution. For a sufficiently large $\rho_c$ the field stays close to $\phi_c$ almost all the way to\footnote{For chameleons the solution only grows in a thin-shell close to the surface, but for large enough densities the field hardly moves at all.} $r=R$. It follows from a second order Taylor expansion around $\phi_S$ that this is guaranteed to be the case as long as
\begin{equation}
\frac{V_{\rm eff,\phi\phi\phi}(\phi_S,\rho_c)(\phi_{\infty}-\phi_S)}{V_{\rm eff,\phi\phi}(\phi_S,\rho_c) m_SR} \ll 1.
\end{equation}
When all these conditions are satisfied, there exists of  a critical solution in the limit $\rho_c \to \infty$ which reads
\begin{eqnarray}
\phi &=& \phi_c~~~~~~~~~~~~~~~~~~~~~~~~~r<R,\\
\phi &=& \phi_{\infty} + \frac{(\phi_c-\phi_{\infty})R}{r}~~~~~~ r > R,
\end{eqnarray}
which, apart from the numerical value of $\phi_{\infty}$ and $\phi_c$, is completely model independent. This critical solution and its implications, for the case of power-law chameleon theories, was discussed in \cite{chameleon}. Another regime which can be described by  exact solutions without having to solve model dependent equations is realised when $\phi_{\infty}\gg \beta_{\infty}\mpl\Phi_N$. In this regime the theory is effectively linear and the solution reads
\begin{align}
\phi &= \phi_{\infty} + \frac{\beta_{\infty}\rho_c R^2}{6\mpl}\left(\frac{r^2}{R^2} - 3\right)~~r<R,\\
\phi &= \phi_{\infty} -  \frac{\beta_{\infty}\rho_c R^3}{3\mpl r}~~~~~~~~~~~~~~~~~~ r > R,
\end{align}
where $\beta_{\infty} = \beta(\phi_{\infty})$. This is the same type of solution as found in Newtonian gravity and the fifth-force-to-gravity ratio on a test mass outside the body is
\begin{align}
\frac{F_{\phi}}{F_G} =  2\beta_{\infty}^2,
\end{align}
while for the critical solution we find
\begin{align}
\frac{F_{\phi}}{F_G} =  2\beta_{\infty}^2 \left(\frac{|\phi_{\infty}-\phi_c|}{2\beta_{\infty}\mpl\Phi_N}\right).
\end{align}
Comparing the two cases we see that the critical solution corresponds to a screened fifth-force given that
\begin{align}\label{cond}
|\phi_{\infty}-\phi_c| \ll 2 \beta_{\infty}\mpl\Phi_N,
\end{align}
which is exactly the screening condition we have found for chameleons, symmetrons and dilatons by solving the field equation explicitly. It is easy to show that the assumptions we started with do hold for these models. The critical solution, which formally only holds in the limit $\rho_c\to\infty$, will be a good approximation for the case of finite $\rho_c$ as long as the screening condition holds by a good margin. As current local gravity experiments give very tight constraints, if one wants to have cosmological signatures i.e. $\beta_{\infty}= \mathcal{O}(1)$, then this will be true in most cases.

For the case where $|\phi_{\infty}-\phi_c| \sim 2\mpl\beta_{\infty}\Phi_N$ we would have to solve the model dependent equation to get accurate solutions. These solutions will interpolate between the two regimes found above, see e.g. \cite{thinshell} for a thorough derivation of chameleon equations in all possible regimes.

\subsection{Cosmological Scalar Field Dynamics}

Here we consider the cosmological evolution of the scalar field $\phi$ in modified gravity models with a minimum of $V_{\rm eff}(\phi)$ at which the scalar field mass $m$ satisfies $m^2 \gg H^2$. The cosmology of the scalar field is  tightly constrained by BBN physics due to the coupling of the scalar field to matter particles. The fact that the scalar field evolves along the minimum of $V_{\rm eff}(\phi)$ implies that the masses of fundamental particles
\be
m_\psi= A(\phi) m_{\rm bare},
\ee
in  which $m_{\rm bare}$ is the bare mass appearing in the matter Lagrangian, evolve too. In practice, tight constraints on the time variation of masses since the time of BBN
\be
\frac{\Delta m_\psi}{m_\psi}= \beta \frac{\Delta \phi}{\mpl},
\ee
where $\Delta \phi$ is the total variation of the field since BBN, impose that $\Delta m_\psi/m_\psi$ must be less than $\sim10\%$. At a redshift of order $z_e\approx 10^9$, electrons decouple and give a "kick" \cite{bbdkw2004}  to the scalar field which would lead to a large violation of the BBN bound. To avoid this, the field must be close to the minimum of $V_{\rm eff}(\phi)$ before $z_e$ and simply  follow the time evolution of the minimum given by
\be
\frac{\rd V}{\rd\phi}\big\vert_{\phi_{\rm min}} = -\beta \frac{\rho_m}{\mpl}.
\ee
Moreover, the total excursion of the scalar field following the minimum must be small enough. In practice, we will always assume that $|\phi/\mpl|\ll 1$ along the minimum trajectory, implying that the BBN bound for the time dependent minimum is always satisfied. The models are then valid provided the electron "kick" does not perturb the minimum too much. We analyse this now.

The background evolution of the scalar field is governed by the homogeneous scalar field equation
\be
\ddot\phi + 3H \dot\phi+ \frac{\rd V_{\rm eff}}{\rd\phi}=0.
\ee
We assume that the contribution of the scalar field to the Hubble rate in the Friedmann equation is negligible until the acceleration of the Universe sets in
\begin{equation}
H^2=\frac{\rho_{\rm rad}+ \rho_m + \rho_\phi}{3\mpl^2},
\end{equation}
where
\begin{equation}
\rho_\phi= \frac{1}{2}\dot\phi^2 + [A(\phi)-1]\rho_m + V(\phi).
\end{equation}
The models that we consider here have a dynamical minimum located at $\phi_{\rm min}(t)$ such that
\begin{equation}
\frac{\rd V_{\rm eff}}{\rd\phi}\big\vert_{\phi_{\rm min}}=0.
\end{equation}
Defining $\delta\phi\equiv\phi-\phi_{\rm min}$, we have for linear perturbations around the minimum
\be
\ddot\delta \phi +3H \dot\delta\phi +m^2 \delta\phi=F,
\label{KGm}
\ee
where
\be
F=-\frac{1}{a^3}\frac{\rd}{\rd t}\left[a^3 \frac{\rd\phi_{\rm min}}{\rd t}\right].
\ee
Using the minimum equation, we find that
\be
\dot \phi_{\rm min}=\frac{3H}{m^2} \beta A \frac{\rho_m}{\mpl},
\ee
and the forcing term is then
\be
F=-\frac{3\rho_{m0}a^{-3}}{\mpl}\frac{\rd}{\rd t}\left[\frac{A\beta H}{m^2}\right].
\ee
We must also take into account the "kicks" that the field receives every time a relativistic species decouples. These "kicks" correspond to the abrupt variation of the trace of the energy momentum tensor of a decoupling species at the transition between the relativistic and non-relativistic regimes. The abrupt change of $T_\mu^\mu$ for the decoupling species happens on a time scale much smaller than one Hubble time and can be modelled out using an "instantaneous kick" approximation \cite{bbdkw2004} where the contribution to the scalar field equation is a $\delta$-function. For kicks at the decoupling times
$t_j$, the source term becomes
\be
F=-\frac{3\rho_0}{\mpl a^3}\frac{\rd}{\rd t}\left[\frac{A\beta H}{m^2}\right]- A\beta \sum_j \kappa_j H_j \mpl\delta(t-t_j),
\ee
where $\kappa_j\approx g_i/g_\star (m_j)\lesssim 1$ depends on the number of relativistic species $g_\star (m_j)$ at time $t_j$ and the number of degrees of freedom of the decoupling species $g_j$.

Let us now go through the different cosmological eras. During inflation, the Hubble rate is nearly constant and the field is nearly constant\footnote{Note the parameterisation $m(a)=m_0a^{-r}$ to be introduced below only applies when the scalar field is sourced by the pressure-less matter, and does not apply to the inflationary era, in which $\phi$ remains nearly constant simply because the density of the inflaton does so.}. Indeed, the trace of the energy momentum tensor is
\be
T\approx -12 H^2\mpl^2,
\ee
in which $\rho_m=-p_m= 3H^2\mpl^2$ is nearly constant in the slow roll approximation. As a result, the source term in the perturbed scalar field equation vanishes, and averaging over the oscillations with the fast period $1/m\ll1/H$ we have
\begin{equation}
\langle\delta \phi^2\rangle \propto a^{-3},
\end{equation}
implying that the field reaches the minimum of the effective potential very rapidly during inflation.

Assuming that reheating is instantaneous and that the field is not displaced during reheating, the field starts in the radiation era at the minimum of the effective potential during inflation. As the minimum has moved to larger values, the field rolls down towards the new minimum,  overshooting and then stopping at a value
\be
\phi_{\rm overshoot} \approx \phi_{\rm inflation}+ \sqrt{6\Omega_\phi^i}\mpl,
\end{equation}
depending on the initial density fraction $\Omega^i_\phi$ in the scalar field \cite{bbdkw2004}. After this the field is in an  undershoot situation where the field is essentially moved according to the kicks
\be
\ddot \phi +3H \dot\phi =- A\beta \sum_j \kappa_j H_j\mpl\delta\left(t-t_j\right).
\ee
Each kick brings the field to smaller values, with a variation
\begin{equation}
\Delta\phi_j= -\beta_j A_j \kappa_j\mpl,
\end{equation}
in the radiation era \cite{bbdkw2004}. Although the details depend on the kicks and the initial energy density of the field, we can assume that after all the kicks before BBN, the field is close to the minimum of $V_{\rm eff}(\phi)$. We will assume that this is the case by $z_{\rm ini}\approx 10^{10}$ where the matter density is equivalent to the one in dense bodies on Earth today. If this were not the case then the field would move by
\begin{equation}
\Delta\phi_e= -\beta_e A_e \kappa_e\mpl,
\end{equation}
when the electron decouples during BBN, and the masses of particles would vary too much during BBN. Note that for the rest of this subsection a subscript $_e$ will be used to denote the value of a quantity at the electron decoupling.

Hence viable models must be such that the scalar field remains in the neighbourhood of the minimum since well before BBN. In this case, the deviation of the field from the minimum
can be easily obtained from
\begin{eqnarray}
\ddot\delta \phi+3H \dot\delta\phi +m^2 \delta\phi &=&-\frac{3\rho_{m0}a^{-3}}{\mpl}\frac{\rd}{\rd t}\left[\frac{A\beta H}{m^2}\right]\nonumber\\
&&-A_e\beta_e \kappa_e H_e\mpl\delta\left(t-t_e\right),
\end{eqnarray}
where we only take into account the electron kick. Defining $\delta\phi= a^{-3/2} \psi$, we find that
\begin{eqnarray}
\ddot \psi+\left[m^2+\frac{9w}{4}H^2\right]\psi &=& -\frac{3\rho_{m0}a^{-3/2}}{\mpl}\frac{\rd}{\rd t}\left[\frac{A\beta H}{m^2}\right]\\
&&-A_e\beta_e \kappa_e H_e a_e^{3/2}\mpl\delta\left(t-t_e\right).\nonumber
\end{eqnarray}
As $m^2\gg H^2$, the solution is obtained using the WKB approximation and reads
\begin{eqnarray}
\frac{\delta\phi}{\mpl} &=& -\frac{9\Omega_{m0} H_0^2}{ a^{3}m^2}\frac{\rd}{\rd t}\left[\frac{A\beta H}{m^2}\right]\\
&& -\Theta\left(t-t_e\right)A_e\beta_e\kappa_e \frac{H_e}{\sqrt{m_em}}\frac{a_e^{3/2}}{a^{3/2}}\sin \int_{t_e}^t m(t') \rd t',\nonumber
\end{eqnarray}
in which the second term is only present when $t>t_e$, $\Theta$ being the Heaviside function. We will always assume that $\beta$ and $m$ vary over cosmological times, hence we have
\be
\frac{\rd}{\rd t}\left[\frac{A\beta H}{m^2}\right]= g(t)\frac{A\beta H^2}{m^2},
\ee
in which $g(t)$ is a slowly-varying function of time whose value is of order unity. Averaging over the rapid oscillations, we have
\begin{eqnarray}
\frac{\langle\delta\phi^2\rangle}{\mpl^2} &=& \frac{81\Omega_{m0}^2 g^2 A^2 \beta^2}{a^6} \frac{H_0^4}{m_0^4} \frac{m_0^4}{m^4}\frac{H^4}{m^4} \nonumber\\
&& + \frac{A_e^2\beta_e^2\kappa_e^2}{2}\frac{a_e^3}{a^3}\frac{H^2_e}{m_e^2}\frac{m_e}{m}.
\end{eqnarray}
The first terms is of order $\beta_0^2 H_0^8/m_0^8 \ll 1$ now, implying that it has a negligible influence on the particle masses. This guarantees that the minimum is indeed a solution of the equations of motion.  The second term corresponds to the response of the scalar field to a kick.
It is initially very small as suppressed by $H_e^2/m_e^2\ll 1$, implying a tiny variation of the fermion masses during BBN. Its influence increases with time as $1/ma^3$ and we must impose that this never compensates the fact that $H_e^2/m_e^2$ is extremely small.

Consider an interesting example with $m(a)=m_0a^{-r}$ which will reappear later. In such a case the second term in the above equation can be rewritten as
\begin{eqnarray}
\frac{A_e^2\beta_e^2\kappa_e^2}{2}\frac{a_e^3}{a^3}\frac{H^2_e}{m_e^2}\frac{m_e}{m} &\sim& \frac{H_0^2}{m_0^2}\frac{\Omega_{r0}}{\Omega_{m0}}a^{r-1}_ea^{r-3},
\end{eqnarray}
where we have assumed $A_e^2\beta_e^2\kappa_e^2\sim\mathcal{O}(1)$ and $\Omega_{r0}\ll\Omega_{m0}$ is the fractional energy density for radiation (photons and massless neutrinos) at present. From this formula we can easily see that
\begin{enumerate}
\item when $r<3$ the minimum of $V_{\rm eff}$ given by the minimum equation is an attractor, because the magnitude of the oscillation decreases in time;
\item assuming that $H_0\sim10^{-3}m_0$ (see below) and $\Omega_{m0}\sim10^{3}\Omega_{r0}$, then today we have $\langle\delta\phi^2\rangle/\mpl^{2}\sim10^{-9}a^{r-1}_e$ which is of order one if $r=0$. Clearly, for $r\lesssim2$ the amplitude of oscillation can be too big ($\sqrt{\langle\delta\phi^2\rangle}\gg\phi_{\rm min}$) at early times;
\item if $r\geq3$ which is the case for $f(R)$ gravity models in which $f(R)\sim R+R_0-R_1(R_\star/R)^n$, $\sqrt{\langle\delta\phi^2\rangle}/\mpl$ increases with time but never becomes significantly large. For example, if $r=3$ then $\sqrt{\langle\delta\phi^2\rangle}/\mpl\sim10^{-15}$ today, which means that, although the minimum of $V_{\rm eff}(\phi)$ is not strictly speaking an attractor, it is extremely stable to kicks and governs the background dynamics of the model.
\end{enumerate}

\subsection{The Equation of State}

We have described how the cosmological constraint from BBN imposes that the scalar field must be at the minimum of the effective potential since BBN. As such the minimum of the effective potential acts as a slowly varying cosmological constant. We have also seen that when $m^2 \gg H^2$, a large class of models are such that the minimum is stable. In this case, the dynamics are completely determined by the minimum equation
\be
\frac{\rd V}{\rd\phi}\big\vert_{\phi_{\rm min}}= -\beta A \frac{\rho_m}{\mpl}.
\end{equation}
In fact, the knowledge of the time evolution of the mass $m$ and the coupling $\beta$ is enough to determine the time evolution of the field.
Indeed, the mass at the minimum of $V_{\rm eff}$,
\be m^2\equiv\frac{\rd^2 V_{\rm eff}(\phi)}{\rd\phi^2}\big\vert_{\phi_{\rm min}},
\ee
and the minimum  relation leads to
\be
V''\equiv \frac{\rd^2V}{\rd\phi^2}= m^2 (a) -  \beta ^2 A(\phi) \frac{\rho_m}{\mpl^2}-\frac{\rd \beta}{\rd\phi} A (\phi) \frac{\rho_m}{\mpl},
\ee
where the couplings to matter $\beta$ can be field dependent. Using the minimum equation, we deduce that the field evolves according to
\begin{eqnarray}\label{eq:min_eqn}
\frac{\rd\phi}{\rd t}=\frac{3H}{m^2} \beta A \frac{\rho_m}{\mpl}.
\end{eqnarray}
This is the time evolution of the scalar field at the background level since the instant when the field starts being at the minimum of the effective potential. In particular, we have
\be
\frac{1}{2}\left(\frac{\rd\phi}{\rd t}\right)^2= \frac{27}{2} \Omega_m \beta^2 A^2\left(\frac{H}{m}\right)^4 \rho_m
\ee
which is tiny compared to $\rho_m$.

Because of the interaction between the scalar field and matter, the energy momentum tensor of the scalar field is not conserved. Only the total energy momentum
\begin{equation}
\dot \rho_{\rm tot} = -3H (\rho_{\rm tot} + p_{\rm tot})
\end{equation}
is conserved, where the total energy density is
\begin{equation}
\rho_{\rm tot} \equiv \rho_m  + \rho_\phi
\end{equation}
with
\begin{eqnarray}
\rho_\phi &=& \frac{\dot \phi^2}{2} + V_{\rm eff}(\phi),\\
\ p_{\rm tot}\equiv p_\phi &=& \frac{\dot \phi^2}{2} - V(\phi),
\end{eqnarray}
and where we have neglected the radiation component in the matter era. It is crucial to notice that the energy density of the scalar field involves the effective potential $V_{\rm eff}$ while the pressure only involves $V$. This is a crucial feature of scalar-tensor theories.

We can define the effective equation of state of the dark energy fluid as
\begin{equation}
w_{\phi}= \frac{p_\phi}{\rho_\phi}.
\end{equation}
Using the Friedmann equation we find the Raychaudhuri equation involving the effective equation of state $w_\phi$ as
\begin{eqnarray}
\frac{\ddot a}{a} &=& -\frac{1}{6\mpl^2}\left[\rho_m+(1+3w_{\phi})\rho_{\phi}\right]\nonumber\\
&\equiv& -\frac{1}{6\mpl^2} (1+3w_{\rm tot}) \rho_{\rm tot}
\end{eqnarray}
where we have defined the total equation of state
\be
w_{\rm tot} = \frac{p_{\rm tot}}{\rho_{\rm tot}}.
\ee
The universe is accelerating provided $\ddot a \ge 0$ which leads to
\begin{equation}
w_{\rm tot}\le-\frac{1}{3}
\end{equation}
as expected,
which is equivalent to
\begin{equation}
w_\phi\le -\frac{1}{3}\left(1+ \frac{\rho_m}{\rho_\phi}\right).
\end{equation}
The situation of the modified gravity models can be easily analysed as
\be
w_\phi +1 = \frac{\dot \phi^2 +(A-1) \rho_m}{\frac{\dot \phi^2}{2} + V(\phi) +(A-1)\rho_m}.
\ee
which can approximated as
\be
w_\phi +1 \approx \frac{\dot \phi^2}{V(\phi)} + (A-1)\frac{\Omega_m}{\Omega_{\phi}}
\ee
The first term corresponds to the usual quintessence contribution and the second term can be approximated as: $\frac{\beta\phi}{\mpl}\frac{\Omega_m}{\Omega_{\phi}} \sim -\frac{\beta}{\mpl}\frac{V_{,\phi}}{V_{,\phi\phi}}\frac{\Omega_m}{\Omega_{\phi}} = 3\beta^2\Omega_m\frac{H^2}{m^2}\frac{\Omega_m}{\Omega_{\phi}}$.
This implies that
\be
w_\phi +1 \approx  (A-1)\frac{\Omega_m}{\Omega_{\phi}} \approx 3 \Omega_m \beta^2\left(\frac{H}{m}\right)^2 \frac{\Omega_m}{\Omega_\phi}.
\ee
In the recent past of the Universe where $\Omega_m$ and $\Omega_\phi$ are of the same order of magnitude, this implies that
the background scalar field acts as a cosmological constant due to the large $H^2/m^2$ suppression. In the past, the background cosmology deviates from a $\Lambda$CDM model only if $\Omega_\phi$ becomes so small that it compensates $m^2/H^2$. We will not consider this situation in the following.

\section{Modified gravity tomography}

\subsection{Reconstruction of the Dynamics}

We have seen that when $m^2 \gg H^2$ a large class of models are such that the minimum of the effective potential is stable or quasi-stable, and in these cases the dynamics are completely determined by the minimum equation
\be
\frac{\rd V}{\rd\phi}\big\vert_{\phi_{min}}= -\beta A \frac{\rho_m}{\mpl}.
\end{equation}
In fact, the knowledge of the time evolution of the mass $m$ and the coupling $\beta$ is enough to determine the bare potential $V(\phi)$ and the coupling function $A(\phi)$ completely. To see this, integrating Eq.~(\ref{eq:min_eqn}) once, we find
\begin{equation}\label{eq:Vofphi}
\phi(a)=  \frac{3}{\mpl}\int_{a_{\rm ini}}^a \frac{\beta (a)}{a m^2(a)}\rho_m(a)\rd a +\phi_c,
\end{equation}
where $\phi_c$ is the initial value of the scalar field at $a_{\rm ini}<a_{\rm BBN}$ and we have taken $A(\phi)\approx1$ as the temporal variation of fermion masses must be very weak. If the coupling $\beta$ is expressed in terms of the field $\phi$ and not the scale factor $a$, this is also equivalent to
\be
\int_{\phi_c}^\phi \frac{\rd\phi}{\beta(\phi)}=  \frac{3}{\mpl}\int_{a_{\rm ini}}^a \frac{1}{a m^2(a)}\rho_m(a)\rd a.
\ee
Similarly the minimum equation implies that the potential can be reconstructed as a function of time
\begin{equation}\label{eq:V}
V=V_0 - \frac{3}{\mpl^2}\int_{a_{\rm ini}}^a \frac{\beta^2(a)}{am^2(a)} \rho_m^2(a) \rd a,
\end{equation}
where $V_0$ is the initial value of the potential at $a=a_{\rm ini}$. This defines the bare scalar field potential $V(\phi)$ parametrically when $\beta (a)$ and $m(a)$ are given. Hence we have found that the full non-linear dynamics of the theory can be recovered from the knowledge of the {\it time} evolutions of the  mass and the coupling to matter since before BBN.

\subsection{Tomography}

The previous reconstruction mapping gives a one-to-one correspondence between the scale factor $a$ and the value of the field $\phi(a)$ in the cosmic background. As the scale factor is in a one-to-one correspondence with the matter energy density $\rho_m(a)$, we have obtained a mapping $\rho_m \to \phi(\rho_m)$  defined using the time evolution of $m(a)$ and $\beta(a)$ only. Given these evolutions, one can reconstruct the dynamics of the scalar field for densities ranging from cosmological to solar system values using Eq.~(\ref{eq:Vofphi}) and Eq.~(\ref{eq:V}). By the same token, the interaction potential can be reconstructed for all values of $\phi$ (and $\rho_m$) of interest, from the solar system and Earth to the cosmological background now: a tomography of modified gravity.

In particular, we can now state the screening condition of modified gravity models as
\be\label{eqq}
\int_{a_{\rm in}}^{a_{\rm out}} \frac{\beta (a)}{a m^2(a)}\rho_m(a)\rd a \ll \beta_{\rm out}\mpl^2\Phi_N,
\ee
with constant matter densities $\rho_{\rm in,out}= \rho_m(a=a_{\rm in,out})$ inside and outside the body respectively, and where we have defined $\beta_{\rm out}\equiv\beta(a=a_{\rm out})$. It is remarkable that the gravitational properties of the screened models are captured by the cosmological mass and coupling functions only.

\subsection{Dilatons}

Let us consider a first example:  the dilaton models in which the coupling function $\beta(\phi)$ vanishes for a certain value $\phi_\star$ of the scalar field $\phi$. On the other hand, we assume that the potential is positive definite and is of runaway type.  It is enough to study the dynamics in the vicinity of the field $\phi_\ast$, where
\be
\beta (\phi) \approx A_2\mpl(\phi-\phi_\star),
\ee
from which we deduce that
\be
\ln\left\vert\frac{\phi-\phi_\star}{\phi_c-\phi_\star}\right\vert = 9 A_2\mpl^2\Omega_{m0}H_0^2 \int_{a_{\rm ini}}^a \frac{\rd a}{a^4 m^2 (a)},
\ee
and therefore
\begin{equation}
\vert \beta (\phi)\vert = \vert \beta (\phi_c)\vert  \exp\left[9 A_2\mpl^2\Omega_{m0}H_0^2 \int_{a_{\rm ini}}^a \frac{\rd a}{a^4 m^2 (a)}\right].
\label{eq:beta}
\end{equation}
In particular, we find the relation between the coupling at the initial time and other cosmological times.

The initial coupling (taken at $a_{\rm ini}< a_{\rm BBN}$) is the same as in dense matter on Earth, {\it as long as} the field minimises its effective potential in a dense environment, and it is related to the cosmological value of $\beta$ today, $\beta(\phi_0)$, by
\begin{equation}
\vert \beta (\phi_0)\vert = \vert \beta (\phi_c)\vert \exp\left[9 A_2\mpl^2\Omega_{m0}H_0^2 \int_{a_{\rm ini}}^1 \frac{\rd a}{a^4 m^2 (a)}\right].
\end{equation}
It is possible to
have a very small coupling in dense matter $\vert \beta (\phi_c)\vert \ll 1$ for any value of the coupling on cosmological scales $\vert \beta (\phi_0)\vert$ provided that $A_2 >0$ and that the time variation of $m(a)$ is slow and does not compensate the $1/a^4$ divergence in the integrand. In this situation, the coupling function $\beta$ converges exponentially fast towards zero: this is the Damour-Polyakov mechanism \cite{dp1994}. The fact that $A_2>0$ guarantees that the  minimum of the coupling function is stable and becomes the minimum of the effective potential which attracts the scalar field in the long time regime. If $A_2<0$, the effect of the coupling is destabilising and implies that $\phi$ diverges exponentially fast away from $\phi_\star$.

Alternatively, a  smooth variation of the coupling function to matter in the cosmological background and therefore interesting consequences for the large-scale structure can be achieved when the evolution of the mass of the scalar field compensates  the $1/a^4$ factor in the radiation era and evolves in the matter era. This is obtained for models with
\be
m^2 (a) = 3 A_2 H^2 (a)\mpl^2.
\ee
Indeed, $H(a) \sim a^{-2}$ in the radiation era, which implies that the time variation of $\beta$ between BBN and matter-radiation equality is
\be
\beta(\phi)=\beta(\phi_c)\exp\left[3\frac{\Omega_{m0}}{\Omega_{r0}}(a- a_{\rm ini})\right],
\ee
and in the matter dominated era
\be
\beta(\phi)= \beta\left(\phi_{\rm eq}\right)\left(\frac{a}{a_{\rm eq}}\right)^{{3}}= \beta (\phi_{\rm eq})\frac{\rho_m\left(a_{\rm eq}\right)}{\rho_m(a)},
\ee
where a subscript $_{\rm eq}$ denotes the value of a quantity at the matter-radiation equality. This is the behaviour of the dilaton models we have already analysed gravitationally in \S~IIB2.

\subsection{Symmetron}

In the symmetron models the coupling to matter vanishes identically in dense regions or at redshifts $z>z_\star$, while a larger coupling is obtained after a transition at a redshift $z_\star$ and in the low matter-density regions. This can be obtained by choosing
\be\label{symc}
\beta(a)=  \beta_\star \sqrt{1-\left(\frac{a_\star}{a}\right)^3},
\ee
for $z<z_\star$ and $\beta=0, \ z>z_\star$. Similarly we choose
\begin{equation}
m(a)=m_\star\sqrt{1-\left(\frac{a_\star}{a}\right)^3}.
\ee
Using the reconstruction mapping, it is straightforward to find that
\begin{equation}
\phi(a)=\phi_\star \sqrt{1-\left(\frac{a_\star}{a}\right)^3},
\end{equation}
for $z<z_\star$ and $\phi=0$ before. The potential for $z<z_\star$ as a function of $a$ can then be reconstructed, using the technique introduced above, as
\be
V(a)=V_0+\frac{\beta_\star^2\rho_\star^2}{2m_\star^2\mpl^2}\left[\left(\frac{a_\star}{a}\right)^6-1\right],
\ee
where
\be
\rho_\star=\frac{\rho_{m0}}{a_\star^3},
\ee
is the matter density at the transition between $\phi(a)=0$ and $\phi(a)>0$. The potential as a function of $\phi$ is then
\be
V(\phi)=V_0 +\frac{\lambda}{4} \phi^4 -\frac{\mu^2}{2} \phi^2,
\ee
where
\be
\phi_\star= \frac{2\beta_\star \rho_\star}{m_\star^2\mpl},
\ee
and
\be
m_\star = \sqrt {2} \mu,\ \
\lambda = \frac{\mu^2}{\phi_\star^2},
\ee
together with
\be
\beta(\phi)= \frac{\beta_\star}{\phi_\star} \phi.
\end{equation}
This completes the reconstruction of the particular symmetron model presented in \cite{symmetron}  from $m(a)$ and $\beta (a)$.

\subsection{Generalised Symmetrons}

With the parametrisation developed in this paper it is easy to create new models (in a more intuitive way than starting with the Lagrangian) by changing  the mass and coupling functions. Here we give a simple example  by generalising the symmetron models.

We start by generalising the coupling function Eq.~(\ref{symc})
\be
\beta(a)=  \beta_\star\left[(1-\left(\frac{a_\star}{a}\right)^3\right]^{1/q},
\ee
for $z<z_\star$ and $\beta=0$ for $z>z_\star$. Similarly we choose
\begin{equation}
m(a)=m_\star\left[1-\left(\frac{a_\star}{a}\right)^3\right]^{1/p},
\ee
where the field evolves as
\be
\phi(a)=  \phi_\star\left[1-\left(\frac{a_\star}{a}\right)^3\right]^{\frac{1}{m-n}},
\ee
where we have defined
\be\label{symm_ind}
m = \frac{2(p-q+pq)}{p-2q+pq},~~~~ n = \frac{2p-2q+pq}{p-2q+pq},
\ee
and where
\be
\phi_\star= \frac{(m-n)\beta_\star \rho_\star}{m_\star^2\mpl}.
\ee
Eventually we find
\be\label{eq:V_symmetron}
V(\phi) = V_0 +
\frac{(m-n)\beta^2_\star\rho^2_\star}{m^2_\star\mpl^2}\left[\frac{1}{m}\left(\frac{\phi}{\phi_\ast}\right)^{m}-\frac{1}{n}\left(\frac{\phi}{\phi_\ast}\right)^{n}\right]
\ee
and
\be
\beta (\phi)= \beta_\star\left(\frac{\phi}{\phi_\star}\right)^{n-1}.
\ee
The indices $m$ and $n$ should be taken to be even integers to keep the potential symmetric around $\phi=0$. The standard symmetron corresponds to the choice $m/2=n=2$.

We can now show explicitly that this generalised symmetron model has the screening property as we did for the original symmetron model in \S~IIB2. Let us consider a spherically dense body of density $\rho_c$ and radius $R$ embedded in a homogeneous background. The field profile inside the body is
\be
\phi= \phi_S \frac{\sinh m_S r}{m_Sr},  ~~~ r<R
\ee
where
\be
m_S^2 \simeq \left(\frac{d\beta(\phi)}{d\phi}\right)_S\frac{\rho_c}{\mpl} = m_\star^2\frac{n-1}{m-n}\frac{\rho_c}{\rho_\star}\left(\frac{\phi_S}{\phi_\star}\right)^{n-2}
\ee
is the scalar field mass at $r=0$, $\phi_S$ the corresponding field value and $\rho_\star$ is as in the symmetron model the critical matter density when the transition of the minimum of $V_{\rm eff}(\phi)$ from $\phi=0$ to $\phi=\pm\phi_\star$ takes place in the cosmological background.

The field outside the body, on scales shorter than the large range $m_\star^{-1}$, is
\be
\phi= \phi_* + \frac{D}{r}, \ r>R
\ee
Matching at $r=R$ gives us the solution
\begin{align}
&\phi_S\cosh(m_S R) = \phi_\star\\
&D =  \phi_\star R\left(\frac{\tanh(m_S R)}{m_SR}-1\right)
\end{align}
The first condition, which determines $\phi_S$, can be written
\be
\frac{\phi_S}{\phi_\star}\cosh\left[\sqrt{\alpha} \left(\frac{\phi_S}{\phi_\star}\right)^{n/2-1}\right] = 1
\ee
where $\alpha = \frac{n-1}{m-n}\frac{\rho_c}{\rho_\star}(m_\star R)^2$. We can change it into a simple equation for $m_S R$
\be\label{m0eq}
(m_SR)^2 \cosh^{n-2}(m_SR) = \alpha
\ee
From these equations we see that when $\alpha \gg 1$ we get $\phi_S \approx 0$, $m_SR\gg 1$ and therefore $D\approx -\phi_\star R$. Note that if $n>2$ the mass vanishes at $\phi=0$, however, this is not a problem for the screening mechanism. Even though a large $\alpha$ pushes the field down towards $\phi = 0$, $m_S$ is still an increasing function of $\alpha$ according to Eq.~(\ref{m0eq}).

The fifth-force on a test mass outside the body is found to be screened as long as
\be\label{gensym}
|\phi_c - \phi_{\infty}| \ll 2\mpl\beta_*\Phi_N
\ee
where $\phi_c = \phi_S \approx 0$ and $\phi_{\infty} = \phi_\star$. This condition is equivalent to $\alpha \gg 1$ and shows that the screening property is present in this model.

Comparing the case $n=2$ with $n>2$ we find that even though $\phi_S/\phi_\star$ is larger in the latter case, the coupling $\beta(\phi_S)$ is smaller as long as we have screening. This means that the force between two test-masses in a dense environment is more screened for larger $n$. Local constraints for the generalised symmetrons are therefore satisfied for (at least) the same range as the standard symmetron: $m_0/H_0 \gtrsim 10^3$.
\begin{figure}
\begin{center}
\includegraphics[scale=0.25]{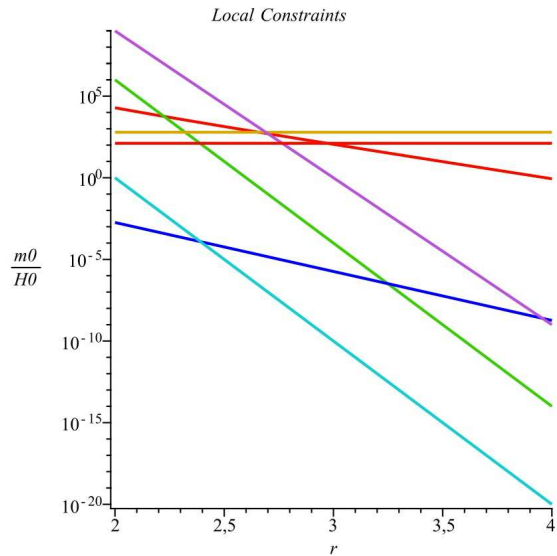}
\caption{The constraints on $m_0/H_0$ as a function of $r$ for $\beta_0=1/\sqrt{6}$ and $s=0$. Valid models must be above the red (solar system), mauve (cavity) green ($m>H$), light blue ($mL\gtrsim 1$), light red ($\dot \mu$) and brown (galaxy) lines. The blue line gives the detectability of effects on the CMB by the  Planck satellite. The strongest constraints are the cavity and galactic bounds for small and large $r$ respectively. Models with $r\gtrsim 3$ satisfy the constraints and can lead to a modified gravity regime on large scales. }
\end{center}
\end{figure}

\section{Reconstructing $f(R)$ models}

\subsection{Gravity Tests and Chameleons}

Consider now the important case of a non-vanishing coupling function $\beta(a)$. Defining $\beta (a)= \beta_0 g(a)$  and $m= m_0 f(a)$, we find that
\begin{equation}\label{eq:phi_chameleon}
\frac{\phi -\phi_c}{\mpl}= 9 \beta_0\Omega_{m0}\frac{H_0^2}{m^2_0} \int_{a_{\rm ini}}^a \rd a\frac{g(a)}{a^4 f^2(a)},
\end{equation}
which allows one to test the screening properties of these models.

Let us first consider the solar system tests. Evaluating Eq.~(\ref{eq:phi_chameleon}) in the galactic background, we find that\footnote{Again, here for simplicity we have assumed that the scalar field minimises $V_{\rm eff}(\phi)$ in the galactic background. While this is true for certain parameter space, in general it should be tested against numerical simulations.}
\be
\frac{\phi_{\rm gal} -\phi_c}{\mpl}= 9 \beta_0\Omega_{m0}\frac{H_0^2}{m^2_0} \int_{ a_{\rm ini}}^{a_{\rm gal}} \rd a \frac{g(a)}{a^4 f^2(a)},
\ee
where $a_{\rm gal}\approx 10^{-2}$ is the scale factor when the matter density in the cosmological background equals the galactic density $\rho_{\rm gal}\approx 10^6 \rho_c$. Defining
\be
\frac{\Delta R}{R}=\frac{\phi_{\rm gal} -\phi_c}{6\mpl\beta_c\Phi_\odot},
\ee
where $R$ is the radius of a spherical body, the modification of gravity in the solar system has a strength
\be
2\beta_{\rm gal}\beta_c\frac{3\Delta R_{\odot}}{R_{\odot}}.
\ee
In this expression $\beta_{\rm gal}$ is the value of the coupling function $\beta(\phi)$ in the galactic background,  $\Phi_\odot$ is the value of the Solar Newtonian potential ($\Phi_\odot\sim 10^{-6}$) and $\beta_c$ is the coupling inside a dense body. The magnitude should be less than $10^{-5}$ to comply with the Cassini bound in the solar system \cite{Bertotti:2003rm}. This condition is independent of $\beta_c$ and reads
\be
\beta_0\beta_{\rm gal} \int_{ a_{\rm ini}}^{a_{\rm gal}} \rd a\frac{g(a)}{a^4 f^2(a)}\lesssim 10^{-5} \frac{m_0^2}{9\Omega_{m0} H_0^2} \Phi_\odot.
\label{cas}
\ee
The integral
\be I\equiv\int_{a_{\rm ini}}^{a_{\rm gal}} \rd a\frac{g(a)}{a^4 f^2(a)},
\ee
is potentially divergent for small values of $a_{\rm ini}\sim10^{-10}$. Hence we must impose that $f(a)^2/g(a)$ compensates the $1/a^4$ divergence in the integrand. As mentioned above, we have assumed that galaxies are screened to minimise the disruption of their dynamics, although the necessity of this condition should be ascertained using $N$-body simulations \cite{hs2007}. Enforcing the screening condition imposes
\be
\vert \phi_{\rm gal} -\phi_0\vert \lesssim 6 \beta_0\mpl \Phi_{\rm gal},
\ee
in which the galactic Newtonian potential is $\Phi_{\rm gal}\sim 10^{-6}$ and
\be
\frac{\phi_0 -\phi_{\rm gal}}{\mpl}= 9 \beta_0\Omega_{m0}\frac{H_0^2}{m^2_0} \int_{ a_{{\rm gal}}}^{1} \rd a\frac{g(a)}{a^4 f^2(a)}.
\ee
A slightly stronger bound is obtained from the Lunar Ranging experiment \cite{Williams:2012nc} with the $10^{-5}$ on the right-hand side of Eq.~(\ref{cas}) replaced by $10^{-7}$.

Strong constraints can also be obtained from laboratory experiments. Using the fact that the initial matter density at $z_{\rm ini} \sim 10^{10}$ is roughly the same as that in a typical test mass in the laboratory, gravity is not modified provided test bodies are screened, {\it i.e.},
\be \vert \phi_{\rm lab}-\phi_c\vert \lesssim 2\beta_c\mpl\Phi_{\rm lab},
\ee
where $\Phi_{\rm lab}\sim 10^{-27}$ for typical test bodies in cavity experiments of size $L$, and $\phi_{\rm lab}= \phi (a_{\rm lab})$ is determined by $m(a_{\rm lab}) \sim 1/L$ (see the appendix for more details).

\subsection{$f(R)$ Gravity Reconstruction}

Viable $f(R)$ models are nothing but chameleons \cite{bbds2008} with a constant value of the coupling function $\beta(\phi)=1/\sqrt{6}$. We have already described the background dynamics of these models. Here we shall derive the mapping between the evolution of the scalar field mass $m(a)$ and the function $f(R)$ for curvature values ranging from the ones in dense bodies to cosmological ones. These models are equivalent to chameleon models where the potential is given by\footnote{In the discussion of $f(R)$ gravity we shall use $R$ to denote the Ricci scalar.}
\be
V(\phi)= \mpl^2 \frac{Rf_R-f}{2f_R^2}
\end{equation}
in which $f_R={\rd f}/{\rd R}$. The mapping between $R$ and $\phi$ is given by
\be
f_R=\exp\left(-2\beta\frac{\phi}{\mpl}\right).
\ee
Given the mass function $m(a)$, we have
\begin{equation}
\phi(a)=  {9\beta\Omega_{m0}H_0^2}{\mpl}\int_{a_{\rm ini}}^a \frac{\rd a}{a^4 m^2(a)}  +\phi_c,\label{phi},
\end{equation}
and
\be
V=V_0 -3  \int_{a_{\rm ini}}^a \frac{\beta^2}{am^2(a)} \frac{\rho_m^2(a)}{\mpl^2}\rd a.
\end{equation}
We can reconstruct $R(a)$ using the fact that
\be
R(\phi)= -e^{2\beta\frac{\phi}{\mpl}} \frac{1}{\beta\mpl}\frac{\rd}{\rd\phi}\left[e^{-4\beta\frac{\phi}{\mpl}}V(\phi)\right],
\end{equation}
and $f(R)$ using
\be
f(R)= R(\phi) e^{-2\beta\frac{\phi}{\mpl}} -\frac{2}{\mpl^2} e^{-4\beta\frac{\phi}{\mpl}}V(\phi),
\ee
which is equivalent to
\be
f(R)= \frac{2}{\mpl^2}e^{-4\beta\frac{\phi}{\mpl}}V(\phi)-\frac{1}{\beta\mpl}e^{-4\beta\frac{\phi}{\mpl}} \frac{\rd V}{\rd\phi},
\ee
once we have obtained $V(\phi)$ from the above implicit parameterisation.

When $\beta\phi/\mpl\ll  1$ as required from the BBN constraints, the above equations can be simplified and read
\be
f(R) = R-2 \frac{V(\phi)}{\mpl^2}
\ee
where
\be
R(\phi)=-\frac{1}{\beta\mpl} \frac{\rd V}{\rd\phi} +\frac{4}{\mpl^2}V(\phi).
\ee
This is the parametric reconstruction mapping of $f(R)$ models.

\subsection{Large Curvature $f(R)$ Models}

We can apply these results to the case with $m=m_0a^{-r}$ leading to models  where
\be
\frac{\phi-\phi_c}{\mpl}= \frac{9\Omega_{m0}\beta H_0^2}{(2r-3)m_0^2}a_{\rm ini}^{2r-3}\left[\left(\frac{a}{a_{\rm ini}}\right)^{2r-3}-1\right],
\end{equation}
which reduces to
\be
\frac{\phi-\phi_c}{\mpl}= \frac{9\Omega_{m0}\beta H_0^2}{(2r-3)m_0^2}a^{2r-3}
\end{equation}
at late times. Similarly we have
\begin{equation}
V(a)=V_0 -\frac{3\beta^2 \rho_{m0}^2}{2(r-3)\mpl^2 m_0^2} \left(a^{2r-6}-a_{\rm ini}^{2r-6}\right).
\end{equation}
Now for late enough times we have
\begin{equation}
V= V_0 -C\left[\frac{\phi-\phi_c}{\mpl}\right]^{\frac{2(r-3)}{2r-3}}
\end{equation}
for a constant $C$. Notice that for $3/2<r<3$, these models are chameleons with an inverse power law potential $V(\phi)\sim \phi^{-n}$ with
\be
n=2\frac{3-r}{3-2r}.
\ee
We can equivalently find that
\begin{equation}
R(\phi)\approx \frac{2C}{\beta\mpl^2}\frac{r-3}{2r-3}\left[\frac{\phi-\phi_c}{\mpl}\right]^{-\frac{3}{2r-3}} + 4\frac{V_0}{\mpl^2}.
\ee
Finally we find that
\be
f(R)= R -\frac{2}{\mpl^2}\left[V_0+C\left(\frac{R-4\frac{V_0}{\mpl^2}}{R_\star}\right)^{-n}\right],
\ee
where $R_\star = 2(r-3) C/\left[(2r-3)\beta\mpl^2\right]$ and
\be
n= \frac{2}{3}(r-3).
\ee
Large curvature models are defined for $r>3$ here. This completes, in this particular example, the reconstruction of the $f(R)$ models from the knowledge of the function $m(a)$.

The gravitational constraints for these models have been fully analysed  in \cite{bdl2012}. We have summarised these constraints in Figure~1 where we see that the strongest constraints on the range of the scalar interaction arise for $r\lesssim 3$, {\it i.e.}, for inverse power law chameleon models. For $r \gtrsim 3$, {\it i.e.}, for large curvature $f(R)$ models, the screening of the Milky Way is a loose constraint which needs to be further analysed with $N$-body simulations.

\subsection{ Comparison with the $B$-parameterisation}

The $f(R)$  theories are generally parameterised using \cite{shs2007}
\be
B=\frac{f_{RR}}{f_R} H \frac{\rd R}{\rd H},
\ee
and $f_R-1$ now. As $\phi/\mpl \ll 1$ we have that
\be
f_R-1= -2\beta\frac{\phi}{\mpl},
\ee
allowing one to reconstruct the field history entirely:
\be
f_R-f_{R0}=  {18\beta^2\Omega_{m0} H_0^2}\int_{a}^1 \frac{1}{a^4 m^2(a)} \rd a ,\label{phi}
\ee
which depends on the mass evolution uniquely. This can be rewritten using the $B$-function. In fact, using
\be
\frac{\rd H}{H}=-\frac{3}{2}(1+w) H\rd t,
\ee
in an era dominated by a fluid of equation of state $w$, we find that
\be
B=-\frac{f_{RR}}{f_R}\frac{2}{3(1+w)} \frac{\dot{R}}{H}.
\ee
With $f_R=e^{-2\beta\frac{\phi}{\mpl}}$
we have
\be
f_{RR} \frac{\rd R}{\rd t}= -2 \beta \frac{f_R}{\mpl} \frac{\rd\phi}{\rd t}
\ee
 and therefore
\be
B= \frac{4 \beta}{3(1+w)\mpl} \frac{\rd\phi}{H\rd t},
\ee
and using the minimum equation we get
\be
B=\frac{6\beta^2}{1+w}\Omega_m \frac{H^2}{m^2}.
\ee
Because $\beta=1/\sqrt{6}$, in the matter dominated era this gives
\begin{equation}
B=\Omega_m\frac{ H^2}{m^2},
\ee
which is completely determined by $m(a)$. Hence we find that
\be
f_R-f_{R0}=  3\int_{a}^1 \frac{B(a)}{a } \rd a.
\ee
The knowledge of $B(a)$ and $f_{R0}$ determines the background evolution in the $f(R)$ gravity models in a completely equivalent way to the $m(a)$ parameterisation.

\section{Growth of Large-scale Structure}

\begin{figure*}
\begin{center}
\includegraphics[scale=0.6]{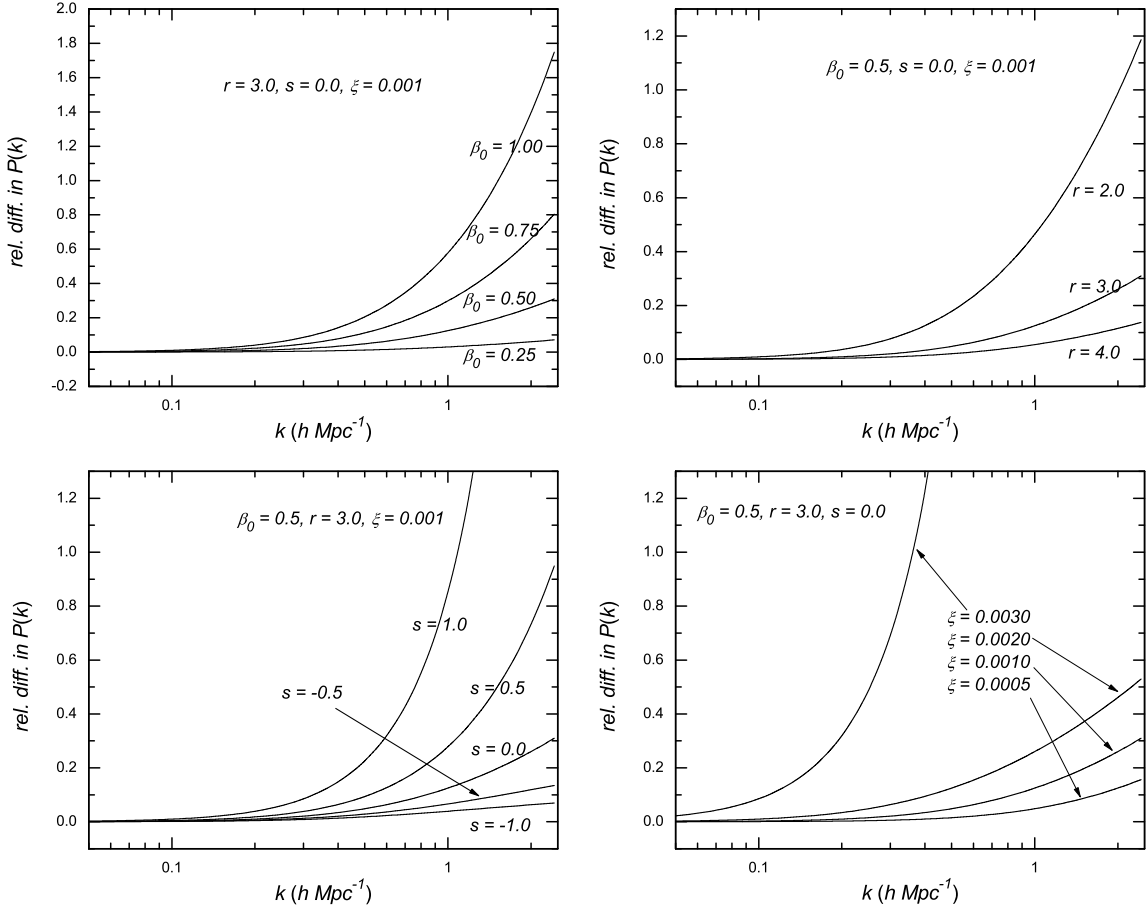}
\caption{The relative difference of the matter power spectrum $P(k)$ in the chameleon model from that in the $\Lambda$CDM model with exactly the same background expansion history, initial conditions and physical parameters. {\it Upper Left panel}: the dependence of the result on the modified gravity parameter $\beta_0$. {\it Upper Right panel}: the dependence of the result on the parameter $r$. {\it Lower Left panel}: the dependence of the result on the parameter $s$. {\it Lower Right panel}: the dependence of the result on the parameter $\xi\equiv H_0/m_0$.}
\label{fig:chameleon}
\end{center}
\end{figure*}

\begin{figure*}
\begin{center}
\includegraphics[scale=0.6]{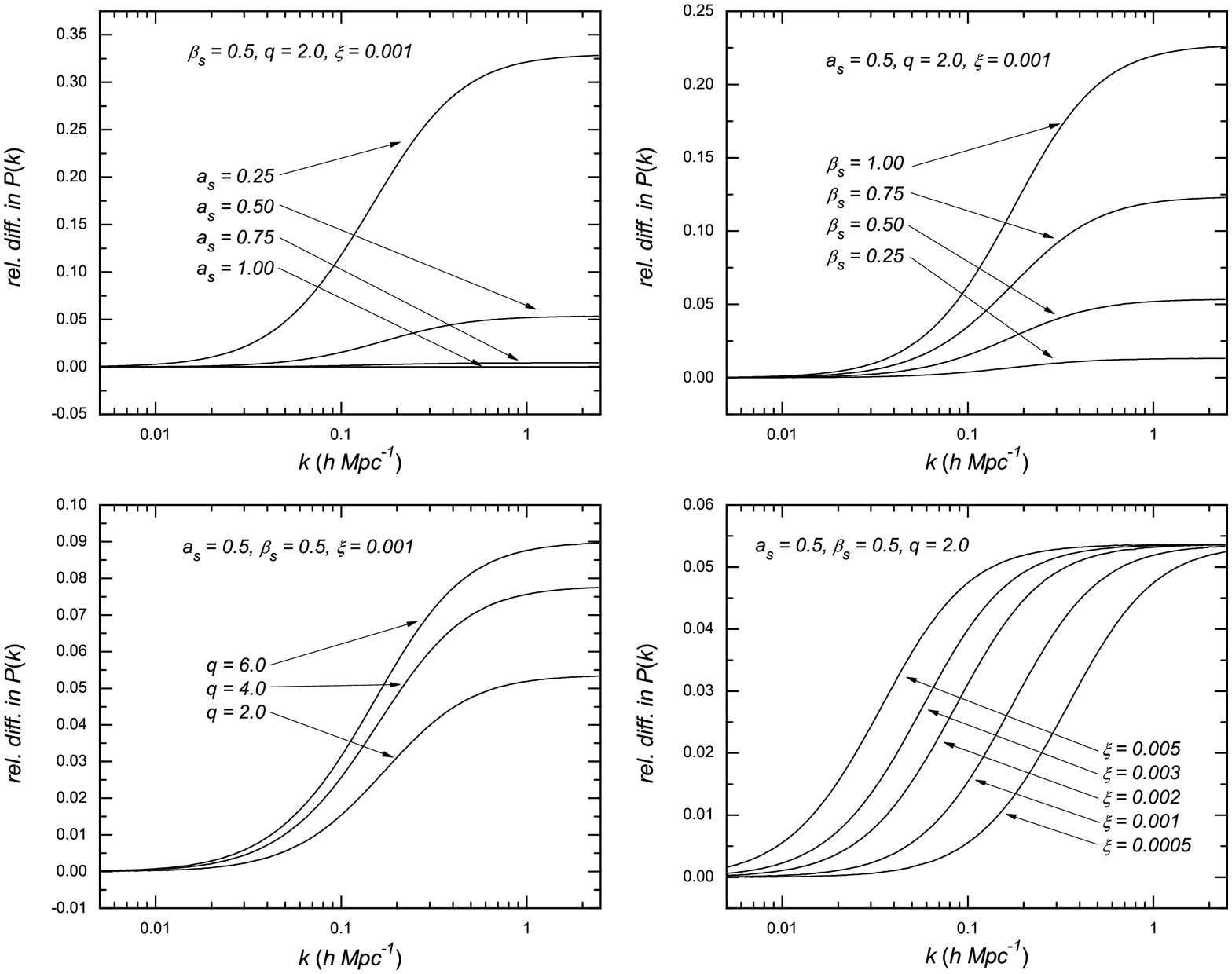}
\caption{The relative difference of the matter power spectrum $P(k)$ in generalised symmetron models from that in the $\Lambda$CDM model with exactly the same background expansion history, initial conditions and physical parameters. {\it Upper Left panel}: the dependence of the result on the parameter $a_\star$ -- the scale factor value at which the symmetry breaking of the effective potential happens. {\it Upper Right panel}:  the dependence of the result on the modified gravity parameter $\beta_\star$. {\it Lower Left panel}: the dependence of the result on the parameter $q$. {\it Lower Right panel}: the dependence of the result on the parameter $\xi\equiv H_0/m_\star$. As an example we have chosen $p=2$.}
\label{fig:symmetron}
\end{center}
\end{figure*}

We have shown that the non-linear structure of the screened models can be reconstructed from the knowledge of the mass and coupling functions. These functions are {\it time dependent} only. In particular, we have seen that this allows one to {\it fully} analysed the gravitational tests and the cosmological background evolution. Moreover we have shown that the cosmological dynamics typically is indistinguishable from a $\Lambda$CDM model at the background level. Here we will find that this is not the case at the perturbative level and that the mass and coupling function allow a full description of the linear and non-linear regimes.

\subsection{Linear Structure Growth}

The linear perturbation equations for a scalar field coupled to matter particles are listed in \cite{lz2009} in the covariant and gauge invariant formalism. Denoting by $\Delta_m$ the density contrast of the pressure-less matter, $v_m$ its velocity and $\delta\phi$ the perturbation\footnote{Note that this is different from above, where we used $\delta\phi$ to denote the oscillation of the background $\phi$ around $\phi_{\rm min}(t)$.} in the scalar field, their evolution equations are as follows:
\begin{eqnarray}
\label{eq:dm_evol}\Delta''_m+\frac{a'}{a}\Delta'_m-\frac{1}{2}\frac{\rho_m}{\mpl^2}a^2\Delta_m\nonumber\\
+k\beta(a)\mpm\left(k\delta\phi-\phi'v_m\right) &=& 0,\\
\label{eq:vm_evol}v'_m+\frac{a'}{a}v_m+\beta(a)\mpm\left(\phi'v_m-k\delta\phi\right) &=& 0,\\
\label{eq:dphi_evol}\delta\phi''+2\frac{a'}{a}\delta\phi'+\left[k^2+a^2m^2(a)\right]\delta\phi\nonumber\\
+\beta(a)\frac{\rho_m}{\mpl}a^2\Delta_m+k\phi'\cal{Z} &=& 0,
\end{eqnarray}
where a prime denotes the derivative with respect to the conformal time, $k{\cal{Z}}=\Psi'$ in the Newtonian gauge  is a variable of the  curvature perturbation which is irrelevant for our discussion since it is multiplied by $\phi'/\mpl\ll {\cal H}= a'/a$, and we have neglected contribution from  radiation as we are focusing on  late times.

Neglecting the terms proportional to $\phi'$ in the above equations we get the following equation \cite{bbdkw2004}
\be
\Delta''_m+\frac{a'}{a}\Delta_m-\frac{1}{2}\frac{\rho_m}{\mpl^2}a^2\Delta_m\left[1+\frac{2\beta^2(a)}{1+\frac{a^2m^2(a)}{k^2}}\right] =0,
\label{eq:dm_evol2}
\ee
where we have used the fact that, given that in Eq.~(\ref{eq:dphi_evol}) the term $k^2+a^2m^2\gg {\cal H}^2$, $\delta\phi$  follows the solution
\begin{equation}\label{eq:dphi}
\delta\phi \approx -\frac{\beta(a)}{k^2+a^2m^2(a)}\frac{\rho_m}{\mpl}a^2\Delta_m,
\end{equation}
and rapidly oscillates around it (see more details below).

On very large scales, $k\ll am(a)$, we can see that Eq.~(\ref{eq:dm_evol2}) reduces to
\begin{eqnarray}
\Delta''_m+\frac{a'}{a}\Delta_m-\frac{1}{2}\frac{\rho_m}{\mpl^2}a^2\Delta_m = 0,
\end{eqnarray}
which governs the growth of matter density perturbation in the $\Lambda$CDM model. The effect of modified gravity is incorporated in the second term in the brackets of Eq.~(\ref{eq:dm_evol2}) and becomes significant when $am(a)/k\lesssim1$, namely for a light scalar field mass $m(a)$ or on small length scales. For all models shown here the CMB spectrum is the same as the $\Lambda$CDM prediction, because the scales relevant for the CMB is very large and therefore not affected by the modified gravity.

In order to illustrate these considerations, we have computed the linear matter power spectra $P(k)$ for a number of generalised chameleon (Fig.~\ref{fig:chameleon}) and symmetron (Fig.~\ref{fig:symmetron}) models.

For the generalised chameleon models, we have used
\be
m=m_0 a^{-r}, \ \beta=\beta_0 a^{-s}
\ee
The impact of gravity tests for $\beta=1/\sqrt{6}$, $s=0$ have been given in Fig.~1. There we can see that values of $r\gtrsim 3$ are favoured by the local gravity tests.
We have varied the four parameters in the parameterisation of $\beta(a)$ and $m(a)$: $\beta_0, r, s$ and $m_0$. Because $m_0$ is not dimensionless, we have defined a new variable $\xi\equiv H_0/m_0$ instead. We find the following results, all as expected:
\begin{enumerate}
\item increasing the coupling $\beta_0$ strengthens the modification of gravity, which causes more matter clustering, resulting in a higher matter power spectrum;
\item $r$ characterises how fast the scalar field mass decreases in time: the higher $r$  the faster it decays. Given that $m_0$ is fixed, a higher value of $r$ means that the Compton wavelength (essentially the range of the modification to gravity) decreases faster in the past, and therefore the modification of gravity starts to take effect later -- this would mean less matter clustering;
\item $s$ specifies how fast the coupling function changes in time: $s=0$ implies $\beta(a)$ remains constant, while $s>0$ ($s<0$) means $\beta(a)$ decreases (increases) in time. If $\beta_0$ is fixed, the larger $s$ is, the larger $\beta(a)$ becomes at high redshifts -- this would mean a stronger modification to gravity and stronger matter clustering;
\item $\xi$ specifies how heavy the scalar field is, or equivalently the range of the modification of gravity: smaller $\xi$ means shorter Compton length of the scalar field, and therefore weaker matter clustering.
\end{enumerate}

The potential of the generalised symmetron models has been given in Eq.~(\ref{symm_ind},\ref{eq:V_symmetron}), but one should be careful that the parameters $p,q$ (or equivalent $n,m$) cannot take arbitrary values. For example, $\phi^{n}$ might not be well-defined if $\phi<0$. Here let us consider the special case with $p=2$ ($n=2$, $m=2+q$), in which the potential becomes
\begin{eqnarray}
V(\phi) &=& V_0+\frac{q\beta^2_\star\rho^2_\star}{m^2_\star\mpl^2}\left[\frac{1}{2+q}\left[\frac{\phi}{\phi_\ast}\right]^{2+q}-\frac{1}{2}\left[\frac{\phi}{\phi_\ast}\right]^{2}\right]
\end{eqnarray}
and this avoids the situation in which the scalar field becomes massless at $\phi=0$. Furthermore, choosing $q=2,4,6,\cdots$ not only ensures that $\phi^{2+q}$ is well-defined for any value of $\phi$, but also makes the potential symmetric about $\phi=0$, as in the original symmetron model. Finally, with $p=2$ another property of the original symmetron model, that $\beta(\phi)\propto\phi$, is preserved as well.

\begin{figure}
\begin{center}
\includegraphics[scale=0.34]{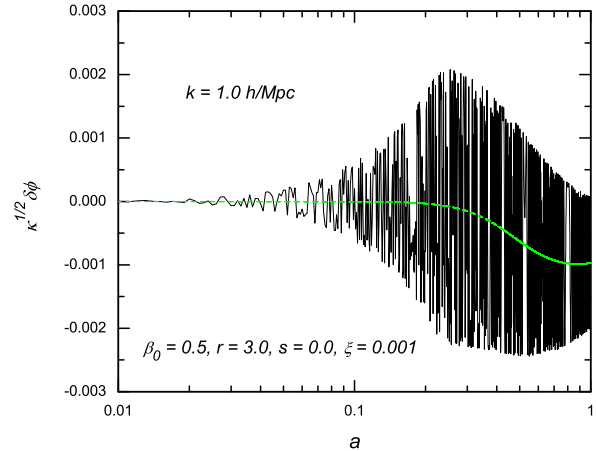}
\caption{An illustration of the time evolution of the scalar field perturbation $\delta\phi$. The black solid curve is the numerical solution while the green dashed curve is the analytical approximation given in Eq.~(\ref{eq:dphi}). The results here are for $k=1h$Mpc$^{-1}$ but the qualitative feature remains for other values of $k$. The modified gravity parameters are shown beside the curves.}
\label{fig:deltaphi}
\end{center}
\end{figure}

Again, the results in Fig.~\ref{fig:symmetron} are as expected:
\begin{enumerate}
\item increasing $a_\star$ implies that the modification of gravity starts to take effect at a later time, and this will weaken the matter clustering;
\item increasing $\beta_\star$ increases the coupling strength overall, and leads to stronger matter clustering;
\item increasing $q$ increases $\beta(a)$ for $a>a_\star$ and causes stronger structure growth;
\item decreasing $\xi$, as in the chameleon case, decreases the range of the modification of gravity, and therefore leads to less matter clustering.
\end{enumerate}

Before we finish this subsection, let us come back to the evolution of the scalar field perturbation $\delta\phi$. As explained above, an analytic approximation to this can be obtained in Eq.~(\ref{eq:dphi}). However, as for the background evolution, where $\phi$ oscillates quickly around $\phi_{\rm min}(t)$, we may expect that the true value of $\delta\phi$ oscillates around the analytic solution as well. This is confirmed in Fig.~\ref{fig:deltaphi}.

In the model shown in Fig.~\ref{fig:deltaphi} we have chosen $r=3.0$. Obviously, the larger $r$ is, the larger the scalar field mass $m(a)$ becomes at early times. A rapid decrease of $m(a)$ would mean that the effective potential for $\delta\phi$ changes its steepness very quickly. Suppose the oscillation of $\delta\phi$ has some initial kinetic energy, then as the effective potential becomes less steep the amplitude of the oscillations increases since the kinetic energy does not disappear quickly.  Consequently, if we increase $r$ further we get even stronger oscillations and if, in contrast, we decrease $r$ then the oscillations become weaker. We have checked explicitly that for $r=1.0$ there is essentially no oscillation.

At late times $H_0/m_0=\xi\sim10^{-3}$, which implies that the period of the oscillation is roughly $10^{-3}$ the Hubble time, and is much longer than the typical time scales for human observations. As a result, one cannot average $\delta\phi$ over several periods to get $\langle\delta\phi\rangle$. Indeed, as the amplitude of oscillation in Fig.~\ref{fig:deltaphi} is bigger than the analytic solution of $\delta\phi$ in Eq.~(\ref{eq:dphi}), the value of $\delta\phi$ one observes at a given time is rather random and could be far from the one given in Eq.~(\ref{eq:dphi}). This is the case for the $f(R)$ gravity model in \cite{hs2007}, where $r=4.5$.

Whilst this seems to be a problem, this is not really the case. Indeed in the solar system the matter density is so high that the oscillation is faster than it is in the cosmological background, and we actually observe the averaged value $\langle\delta\phi\rangle$ . On linear scales, as $\delta\phi$ oscillates, overshooting and undershooting the value given in Eq.~(\ref{eq:dphi}), we have checked by replacing the numerical solution of $\delta\phi$ by the analytical formula given in Eq.~(\ref{eq:dphi}) that we obtain  identical power spectra $P(k)$ in the two approaches. Hence the mean value solution Eq.~(\ref{eq:dphi}) gives a very good description of the statistical properties of linear perturbations.

\subsection{The Jordan Frame Picture}

\begin{figure}
\begin{center}
\includegraphics[scale=0.34]{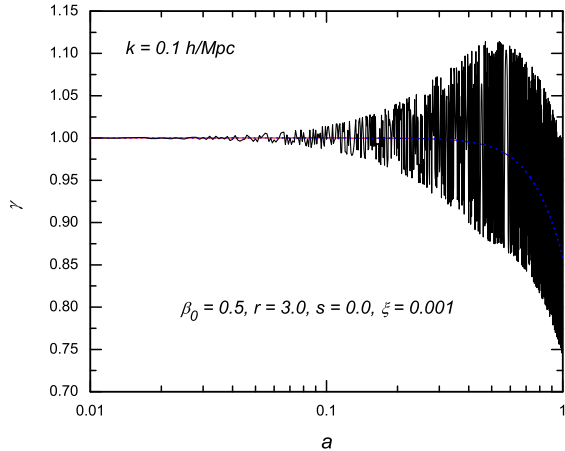}
\caption{The time evolution of $\gamma(k,a)$ for a chosen value of $k=0.1h$Mpc$^{-1}$ as an illustration. The black solid is the full numerical solution, the red dashed curve is obtained using the numerical value of $\Phi_N$ using the analytical solution of $\delta\phi$ given in Eq.~(\ref{eq:dphi}), while the blue solid curve is Eq.~(\ref{eq:linper}). The modified gravity parameters are shown beside the curves.}
\label{fig:gamma}
\end{center}
\end{figure}

In this section we compare our results with a simple and effective way of parameterising linear perturbations which has been used in the literature in the past few years \cite{jz2008,bz2008,aks2008,sk2009,bt2010,dlsccll2010,pskz2010,zllkbz2011} (other interesting and more general approaches for the linear regime include the parameterised Post-Friedmann framework of \cite{bertschinger2006,hs2007b} and the fully covariant parameterisation of \cite{skordis2009,fs2010,zbfs2011}). Such a way of parameterising any modification of  gravity utilises two arbitrary functions $\mu(k,a)$ and $\gamma(k,a)$ through the (modified) Poisson equation
\begin{equation}
-k^2\Psi=4\pi\mu(k,a) G_Na^2\delta\rho_m,
\end{equation}
and the slip relation
\begin{equation}
\Phi=\gamma (k,a) \Psi.
\end{equation}
Here $G_N$ is the bare Newton constant, and $\Psi$ and $ \Phi$ are the two gravitational potentials in the Newtonian gauge:
\begin{equation}
\rd\tilde{s}^2=-a^2(1+2\Psi)\rd\eta^2+a^2(1-2\Phi)\rd x^2,
\end{equation}
in which $(\eta, x)$ are  the conformal time and comoving coordinates.

So far we have focused on the Einstein frame. In the Jordan frame as described by the line element above, the perturbative dynamics can be described using two Newtonian potentials
where we have the relation
\be
\rd\tilde s^2= A^2(\phi) \rd s^2,
\ee
and $ds^2$ is the line element in the Einstein frame expressed in the Newtonian gauge. Expanding in perturbation around a background value with $A[\phi (t)]\approx 1$, we can relate these two potentials to the Einstein frame Newton potential
\begin{eqnarray}\label{eq:PsiPhi}
\Psi &=& \Phi_N +\beta\frac{\delta\phi}{\mpl},\nonumber\\
\Phi &=& \Phi_N- \beta\frac{\delta\phi}{\mpl}.
\end{eqnarray}
Hence we see that in the Jordan frame the two Newtonian potentials are not equal, a fact which can be interpreted as resulting from the existence of a non-anisotropic stress contribution coming from the scalar field. It is useful to define
\be
\epsilon(k,a)= \frac{2\beta^2}{1+\frac{m^2a^2}{k^2}}
\ee
Using the definitions in Eq.~(\ref{eq:PsiPhi}), the analytical approximation for $\delta\phi$ in Eq.~(\ref{eq:dphi}) and the Poisson equation
\begin{eqnarray}
-k^2\Phi_N = \frac{1}{2}\frac{\rho_m}{\mpl^2}a^2\Delta_m,
\end{eqnarray}
it can be derived easily that
\begin{eqnarray}
\gamma (k,a) \equiv \frac{\Phi}{\Psi} &=&\frac{1-\epsilon(k,a)}{1+\epsilon(k,a)},\nonumber\\
\mu(k,a) &=& 1+\epsilon(k,a).
\end{eqnarray}
These results are valid for all the models which can be described by a field tracking the minimum of the effective potential since before BBN. More precisely we find that
\begin{eqnarray}\label{eq:linper}
\mu(k,a) &=& \frac{(1+2\beta^2) k^2 + m^2 a^2}{k^2 +m^2a^2},\nonumber\\
\gamma(k,a) &=& \frac{(1-2\beta^2) k^2 + m^2 a^2}{(1+2\beta^2)k^2 +m^2a^2}.
\end{eqnarray}
These are closely related to the popular parameterisation of modified gravity used in the literature. Here they are valid for any model of modified gravity at the linear level of cosmological perturbations as long as the background cosmology is described by a scalar field slowly evolving in time and following the time dependent minimum of the effective potential where $m^2\gg H^2$.

As a numerical illustration, in Fig.~\ref{fig:gamma} we have compared the function $\gamma(a,k)$ calculated using three different methods: (1) the full numerical solution as shown by the black solid curve, (2) the value obtained by using the definitions in Eq.~(\ref{eq:PsiPhi}), the analytical approximation for $\delta\phi$ in Eq.~(\ref{eq:dphi}) and $\Phi_N$ solved from the Poisson equation numerically (the red dashed curve) and (3) Eq.~(\ref{eq:linper}) as shown by the blue dotted curve. We can see that the latter two agree with each other very well, showing that the parameterisation given in Eq.~(\ref{eq:linper}) works very well in practice and describes the statistical properties of linear perturbations.

The full numerical solution, however, again shows the oscillating behaviour, but the oscillation always centres around the averaged value defined by the previous formulae. As discussed earlier, over many oscillations there will be a cancellation and the net effect on a statistical observable today are the same for all the three curves.

\subsection{$f(R)$ Gravity in the Jordan Frame}

Let us concentrate now on the case of $f(R)$ gravity. The perturbations are then determined by
\begin{eqnarray}
\mu(k,a) &=& \frac{\frac{4}{3} k^2 + m^2 a^2}{k^2 +m^2a^2},\nonumber\\
\gamma(k,a) &=& \frac{\frac{2}{3} k^2 + m^2 a^2}{\frac{4}{3}k^2 +m^2a^2}.
\end{eqnarray}

For large curvature models with $m=m_0 a^{-r}$, this becomes
\begin{eqnarray}
\mu(k,a) &=& \frac{\frac{4}{3} \frac{k^2}{m_0^2}a^{3n+4}  + 1}{\frac{k^2}{m_0^2}a^{3n+4} +1},\nonumber\\
\gamma(k,a) &=& \frac{\frac{2}{3} \frac{k^2}{m_0^2}a^{3n+4}  + 1}{\frac{4}{3}\frac{k^2}{m_0^2}a^{3n+4} +1}.
\end{eqnarray}
When $n = \frac{2}{3}(r-3)\ll 1$, we retrieve the phenomenological parameterisation \cite{bz2008}
\begin{eqnarray}
\mu(k,a) &\approx& \frac{\frac{4}{3} \frac{k^2}{m_0^2}a^{4}  + 1}{\frac{k^2}{m_0^2}a^{4} +1},\nonumber\\
\gamma(k,a) &=& \frac{\frac{2}{3} \frac{k^2}{m_0^2}a^{4}  + 1}{\frac{4}{3}\frac{k^2}{m_0^2}a^{4} +1}.
\end{eqnarray}
Our parameterisation Eq.~(\ref{eq:linper}) covers all the possible $f(R)$ models.

\subsection{Non-linear Effects}

Matter clustering on galactic and cluster scales is an important probe of modified gravity. The nonlinearity in both the structure formation process and the dynamics of the scalar field for scales $k\gtrsim 0.1\ h{\rm Mpc}^{-1}$ require full numerical simulations \cite{lb2011,lztk2011}.

The $\beta(a), m(a)$ parameterisation can completely specify the nonlinear dynamics of $\phi$ with two temporal functions. Indeed, as we have seen above, one can reconstruct the potential $V(\phi)$ and the coupling function together with the background evolution $\phi(a)$. Then one can study the non-linear evolution of the scalar field perturbation which, in the quasi-static limit, are governed by
\be
\nabla^2\phi = \left[\beta(\phi)\frac{\rho_m}{\mpl}-\beta(\overline{\phi})\frac{\overline{\rho}_m}{\mpl}\right]+\frac{\rd V(\phi)}{\rd\phi}-\frac{\rd V(\bar{\phi})}{\rd\phi},
\ee
where the overbar means the background value.


One can easily obtain $\rd V(\phi)/\rd\phi$ analytically or numerically, and this can be used to solve the quasi-static dynamics  numerically. An advantage is that {\it temporal} functions $m(a), \beta(a)$ completely specify the dynamics of $\phi$, in particular its {\it spatial} configuration, and there is no need for a  $k$-space parametrisation.

On linear scales, this is equivalent to the Jordan-frame description with the two spatially dependent function $\mu(k,a)$ and $\gamma(k,a)$ being defined by $\epsilon (k,a)$ which depends on the two functions $m(a)$ and $\beta(a)$, as given in Eq.~(\ref{eq:linper}). But in practice, working with two temporal functions is much more direct. Furthermore, the parameterisation described in Eq.~(\ref{eq:linper}) fails to faithfully describe the nonlinear effects or the environmental dependence. In essence, by going from $m(a)$ and $\beta(a)$ to $\mu(k,a)$ and $\gamma(k,a)$, one not only introduces spatial dependence but also loses the ability to describe nonlinear structure formation: in this sense, we may describe the approach using $\mu(k,a)$ and $\gamma(k,a)$ as the {\it linear parameterisation} of structure formation while  $m(a)$ and $\beta(a)$ provide a  {\it fully non-linear parameterisation} of modified gravity\footnote{Our parameterisation also provides a clear characterisation of the class of physical models (namely a scalar field coupled to matter) considered here, which is important in parameterising modified gravity \cite{zbfs2011}, and not automatically incorporated in the $(\mu,\gamma)$ parameterisation.}.

Past experience has shown that in modified gravity ({\it e.g.}, chameleon and $f(R)$) models, nonlinear effects become important as soon as the linear perturbation result deviates from the corresponding $\Lambda$CDM prediction. This emphasises the importance of using full numerical simulations in the study of these models. However, the full numerical simulations are generally very time and resource-consuming, and are therefore left for future work.

\section{Variation of Constants}

We have seen that the background evolution of the scalar field is specified by the time dependent mass and coupling functions. As the scalar field evolves, the particle masses and the gauge coupling constants change in time too. The time variation of masses and gauge couplings is tightly constrained by laboratory experiments \cite{Luo:2011cf}. In this section, we analyse the time drift of the fine structure constant and the electron to proton mass  ratio.

\subsection{The fine structure constant}

The scalar field  also has an effect  on gauge couplings and particle masses. The fermion masses are given by
\be
m_F(\phi)=A(\phi) m_{\rm bare}.
\ee
where $m_{\rm bare}$ is the bare mass in the Lagrangian. Meanwhile, quantum effects such as the presence of heavy fermions lead to the potential coupling of $\phi$ to photons \cite{Brax:2010uq}
\be
S_{\rm gauge}= -\frac{1}{4g_{\rm bare}^2} \int {\rm d}^4 x \sqrt{-g_{}} B_F(\phi) F_{\mu\nu} F^{\mu\nu},
\ee
where $g_{\rm bare}$ is the bare coupling constant and
\be
B_F(\phi)= 1 + \beta_\gamma\frac{\phi}{\mpl} + \dots.
\ee
The scalar coupling to the electromagnetic field would lead to a dependence of the fine structure constant on $\phi$ as
\be
\frac{1}{\alpha}= \frac{1}{\alpha_{\rm bare}} B_F (\phi),
\ee
implying that
\be
\frac{\dot \alpha}{\alpha}\approx -\beta_\gamma\frac{\dot{\phi}}{\mpl}
\ee
where we have assumed that $\beta_\gamma\phi/\mpl  \ll 1$. Using the evolution equation we find that \be \frac{\dot \alpha}{H \alpha}\approx -9\beta_\gamma \beta \Omega_m \frac{H^2}{m^2}.\ee
Hence the negative variation of the fine structure constant in one Hubble time is related to the small ratio $H/m \ll 1$ and the couplings of $\phi$ to matter and photons.
The best experimental bound on the variation of $\alpha$ now comes from Aluminium and Mercury single-ion clocks \cite{Uzan:2010pm}: $\frac{\dot \alpha}{\alpha}{\large\vert}_0 = (-1.6\pm 2.3)\cdot10^{-17}{\rm yr}^{-1}$. Taking $H_0^{-1} \sim 1.5\cdot10^{10} {\rm yr}$, we get the conservative bound $\left\vert\frac{\dot \alpha}{H\alpha}\right\vert_0  \lesssim 2\cdot 10^{-7}$. As a result, the experimental bounds on the time variation of $\alpha$ lead to constraints on $\beta_0\beta_{\gamma 0}$ as $\beta_0 \beta_{\gamma 0}\lesssim 0.8\cdot10^{-7} \frac{m_0^2}{H_0^2}$. For models with $\beta_0={\cal O}(1)$, $\Omega_{m0}\sim 0.25$ and $m_0/H_0\approx 10^{3}$ where effects on large scale structure are present, $\beta_{\gamma0} \lesssim 0.1$, which is a much tighter bound than present experimental ones $\beta_{\gamma 0} \lesssim 10^{11}$ \cite{chase}.

The time evolution in the past is also particularly interesting. For symmetron models, we find that the time variation of $\alpha$ is
\be
\frac{\dot \alpha}{H \alpha}\approx -9\beta_\star\beta_{\gamma}\Omega_{m0}\left(\frac{H_0}{m_\star}\right)^2 \frac{1}{a^3\sqrt{1-(\frac{a_\star}{a})^3}}
\ee
Here, the time variation of $\alpha$ increases as one reaches the transition $a_\star$. This is a large variation which may happen in the recent past of the Universe and may have observable consequences in the emission lines of distant objects.

It should however be noted that even though $\dot{\alpha}/\alpha$ can be very large, the relative difference of $\alpha$ between the earth and some other sparser place in the Universe is constrained to be less than
\be
\left|\frac{\Delta\alpha}{\alpha}\right| < \frac{\phi_\star\beta_{\gamma}}{\mpl} = 3\beta_\star\beta_{\gamma}\Omega_{m0}\left(\frac{H_0}{m_\star}\right)^2\frac{\rho_\star}{\rho_{m0}}
\ee
If we instead consider a quadratic coupling to photons, $B_F(\phi) = 1 + \frac{A_2^\gamma}{2} \phi^2$, we find
\be
\left|\frac{\Delta\alpha}{\alpha}\right| < A_2^{\gamma}\frac{\phi_\star^2}{2} = 3\beta_\star\beta_{\gamma}\Omega_{m0}\left(\frac{H_0}{m_\star}\right)^2\frac{\rho_\star}{\rho_{m0}}
\ee
where $\beta_{\gamma} = \beta A_2^{\gamma}/A_2 =\phi_\star \mpl A_2^{\gamma} $.

Interestingly, for both cases and for our fiducial parameter values $m_\star \sim 10^3 H_0$, $\rho_\star \sim \rho_{m0}$ and $\beta\sim \beta_{\gamma}  = \mathcal{O}(1)$ this term is of the same order as the claimed variation of $\alpha$ reported in \cite{webb}.

\subsection{The variation of masses}

Fundamental fermions such as the electrons have a universal mass dependence $m_F= A(\phi) m_{\rm bare}$, implying that \be \frac{\dot m_F}{H m_F} = 9\beta^2 \Omega_m \frac{H^2}{m^2}.\ee Nucleons such as the proton have a mass given by the phenomenological formula
\be
m_p=C_{\rm QCD} \Lambda_{\rm QCD} + b_u m_u + b_d m_d + C_p \alpha,
\ee
where $\Lambda_{\rm QCD} \sim 217 {\rm MeV}$ is the QCD scale, $b_u+ b_d \sim 6$, $b_u-b_d \sim 0.5$, $C_{\rm QCD}\sim 5.2$, $m_u^{\rm bare}\sim 5 {\rm MeV}$, $ m_d^{\rm bare}\sim 10 {\rm MeV}$ and $C_p\alpha_{\rm bare} \sim 0.62 {\rm MeV}.$
Assuming conservatively that $\Lambda_{\rm QCD}$ is scalar independent, we get
\be
\frac{\dot m_p}{H m_p} \approx 9\Omega_m \beta \frac{H^2}{m^2}\left ( \frac{b_u m_u^{\rm bare} + b_d m_d^{\rm bare}}{m_p}\beta - \frac{C_p \alpha_{\rm bare}}{m_p} \beta_\gamma\right).
\ee
It is particularly important to study the variation of \be \nu= \frac{m_e}{m_p}\ee from which we find that its time variation  is positive for modified gravity models: \be\frac{\dot \nu}{\nu}\approx 9\Omega_m \beta \frac{H^2}{m^2}\left ( \beta+ \frac{C_p \alpha_{\rm bare}}{m_p} \beta_\gamma\right).\ee The current experimental constraint is $\frac{\dot \nu}{\nu}\large\vert_0 = (-3.8\pm 5.6) 10^{-14}{\rm yr}^{-1}$
which yields the upper bound on $\beta_0$: $\beta_0^2 \lesssim 10^{-5}\frac{m_0^2}{H_0^2}$. For $\beta_0={\cal O}(1)$, this entails that $m_0/H_0\gtrsim 10^{2.5}$. Again for symmetron models, the electron to proton mass ratio would vary rapidly in time around the transition time $a_\star$. It would be interesting to study if such a variation could have relevant effects on the physics of distant objects.

\section{Conclusion}

We have developed a novel parametrisation of modified gravity models first presented in \cite{bdl2012}.
Starting with the time-evolution of the mass and the matter coupling of a scalar field in the cosmological background, we have been able to reverse engineer the complete dynamics of these models in a simple way.

We have applied these results to well-known modified gravity models: chameleons, $f(R)$ gravity, dilatons and symmetrons. In each case, we have explicitly given the mapping and the full reconstruction. We have also shown  how one can apply local constraints using this formalism and then use it to make predictions for linear cosmological perturbations.

New classes of models can be engineered in a more intuitive way than  starting from a Lagrangian. The Lagrangian itself can be completely reconstructed. One only needs to specify two functions whose physical meaning is easily grasped: namely the mass (the inverse range of the fifth-force) and the coupling to matter.

The real strength of this approach compared to existing parameterisations in the literature is that we can reconstruct the whole theory at the linear and non-linear levels and be sure that it corresponds to a concrete physical model defined via a Lagrangian. This effectively supersed existing parameterisations of modified gravity with a screening mechanism by being able to make predictions for non-linear clustering of matter via $N$-body simulations. This will be the subject of future work.

\section*{Acknowledgments}

A.C.D. is supported in part by STFC. B.L. is supported by the Royal Astronomical Society and Durham University. H.A.W. thanks the Research Council of Norway FRINAT grant 197251/V30. P.B. and H.A.W. thanks DAMPT at Cambridge University and H.A.W. thanks IPhT CEA Saclay for hospitality where part of this work was carried out.

\appendix

\section{The Cavity Constraint}

In this appendix, we will explicitly develop the calculation leading to the cavity constraint for chameleon and $f(R)$ models.

Consider a cavity of radius $L$ with a residual density $\rho_{\rm cav}\ll \rho_c$ where $\rho_c$ is the density of the bore surrounding the cavity. The field inside the cavity is $\phi_{\rm cav}$ and deviates slightly from this value across the cavity. Expanding the effective potential around $\phi_{\rm cav}$ and putting $\delta \phi=\phi-\phi_{\rm cav}$, we have
\be
\frac{1}{r^2} \frac{\rd}{\rd r}\left(r^2\frac{\rd}{\rd r}\delta\phi\right)- m_{\rm cav}^2 \delta\phi= V_{{\rm eff},\phi}\left(\phi_{\rm cav}\right)
\ee
where
$m_{\rm cav}$ is the scalar field mass inside the cavity and $ V_{\rm eff,\phi}\equiv\rd V_{\rm eff}(\phi)/\rd\phi$ is nonzero unless $\phi_{\rm cav}$ minimises the effective potential. Inside the cavity the solution is
\be
\delta \phi= A \frac{\sinh (m_{\rm cav}r)}{r} -\frac{ V_{{\rm eff},\phi}\left(\phi_{\rm cav}\right)}{m_{\rm cav}^2};
\ee
outside the cavity we have
\be
\phi= \phi_c + B \frac{e^{-m_c r}}{r},
\ee
where $A, B$ are constants of integral, $\phi_c$ is the minimum of the effective potential outside the cavity and $m_c$ the mass at that minimum.
Matching at $r=L$, we find that
\be
B= \frac{e^{m_cL}}{1+m_cL}\left[\sinh\left(m_{\rm cav}L\right)-m_{\rm cav}L\right]A,
\ee
 and
\begin{eqnarray}
&&A\left[\frac{m_c}{1+m_cL} \sinh\left(m_{\rm cav}L\right) + \frac{m_{\rm cav}}{1+m_cL}\right]\nonumber\\
&=& \phi_c -\phi_{\rm cav} + \frac{V_{{\rm eff},\phi}\left(\phi_{\rm cav}\right)}{m_{\rm cav}^2}.
\end{eqnarray}
Evaluating the solution at the origin and  putting $\delta\phi (r=0)=0$ we have
\be
A=\frac{V_{{\rm eff},\phi}\left(\phi_{\rm cav}\right)}{m_{\rm cav}^3}.
\ee
This leads to
\be
1+ \frac{\sinh\left(m_{\rm cav}L\right)}{m_{\rm cav}L}= -\frac{\phi_{\rm cav}m_{\rm cav}^2}{V_{{\rm eff}, \phi}\left(\phi_{\rm cav}\right)},
\ee
where we have used $m_c L\gg 1$.

For potentials $V\sim 1/\phi^n$ and as long as $\phi_{\rm cav}$ is much less than the effective minimum in the cavity we have
\be
\frac{\sinh\left(m_{\rm cav}L\right)}{m_{\rm cav}L} = n,
\ee
which implies that
\be
m_{\rm cav}L={\cal O} (1),
\ee
where $m_{\rm cav}$ is dominated by the potential term.

\end{document}